\documentclass[structabstract]{aa}  
\usepackage{natbib}
 \bibpunct{(}{)}{;}{a}{}{,} 
 
\usepackage{graphicx}

\usepackage{txfonts}
\def\kms{km s$^{-1}$\space}
\def\kmsnospace{km s$^{-1}$}
\def\micron{$\mu$m\space}

\def\arcsec{$^{\prime\prime}$\space}

\def\h2{H$_2$}

\def\c2{[C\,{\sc ii}]}
\def\13co{$^{13}$CO}
\def\c18o{C$^{18}$O}
\def\12co{$^{12}$CO}

\def\c+{C$^+$}

\def\h2{H$_2$}

\begin{document}
     \title{A {\it Herschel} [C\,{\sc ii}]  Galactic plane survey II:  CO--dark H$_2$ in clouds}
   
\titlerunning{Galactic plane [C\,{\sc ii}] survey of CO--dark H$_2$}
\authorrunning{Langer, Velusamy, Pineda, Willacy, Goldsmith}

   \author{W. D. Langer,
          T. Velusamy,
           J. L. Pineda,
           K. Willacy,
           \and
                  P. F. Goldsmith\thanks{{\it Herschel} is an ESA space observatory with science instruments provided by European-led Principal Investigator consortia and with important participation from NASA. This manuscript is to be published in Astronomy \& Astrophysics in 2014 and is reproduced with permission, {\copyright} ESO.}
 }
          

   \institute{Jet Propulsion Laboratory, California Institute of Technology,
              4800 Oak Grove Drive, Pasadena, CA 91109-8099, USA\\
              \email{William.Langer@jpl.nasa.gov}
             }

   \date{Received 31 July 2013; accepted 20 November 2013}

 

\abstract
{H\,{\sc i} and CO large scale surveys of the Milky Way trace the diffuse atomic clouds and the dense shielded regions of molecular hydrogen clouds, respectively. However, until recently, we have not had spectrally resolved C$^+$ surveys in sufficient lines of sight to characterize the ionized and photon dominated components of the interstellar medium, in particular, the H$_2$ gas without CO, referred to as CO--dark H$_2$, in a large sample of interstellar clouds. }
{To use a  sparse Galactic plane survey of the 1.9 THz (158 $\mu$m) [C\,{\sc ii}] spectral line from the {\textit {Herschel}}  Open Time Key Programme, Galactic Observations of Terahertz C+ (GOT C+), to characterize the H$_2$ gas without CO in a statistically significant sample of interstellar clouds.} 
{We identify individual clouds in the inner Galaxy by fitting the [C\,{\sc ii}]  and CO isotopologue spectra along each line of sight. 
We then combine these spectra with those of H\,{\sc i} and use them along with excitation models and cloud models of C$^+$  to determine the column densities  and fractional mass of CO--dark H$_2$ clouds.}
{We identify1804 narrow velocity [C\,{\sc ii}] components corresponding to interstellar clouds in different categories and evolutionary states.  About 840 are diffuse molecular clouds with no CO,  $\sim$510  are transition clouds containing [C\,{\sc ii}] and $^{12}$CO, but no $^{13}$CO, and the remainder are dense molecular clouds containing $^{13}$CO emission.  The CO--dark H$_2$ clouds are concentrated between Galactic radii of $\sim$ 3.5 to 7.5 kpc and the column density of the CO--dark H$_2$ layer varies significantly from cloud-to-cloud with a global average of 9$\times$10$^{20}$ cm$^{-2}$. These clouds contain a significant fraction by mass of CO--dark H$_2$, that varies from $\sim$75$\%$ for diffuse molecular clouds to $\sim$20$\%$ for dense molecular clouds.}
{We find a significant fraction of the warm molecular ISM gas is invisible in H\,{\sc i} and CO, but is detected in [C\,{\sc ii}]. The fraction of CO--dark H$_2$ is greatest in the diffuse clouds and decreases with increasing total column density, and is lowest in the massive clouds.  The column densities and mass fraction of CO-dark H$_2$ is less than predicted by models of diffuse molecular clouds using solar metallicity, which is not surprising as most of our detections are in Galactic regions where the metallicity is larger and shielding more effective.  There is an overall trend towards a higher fraction of CO--dark H$_2$ in clouds with increasing Galactic radius, consistent with lower metallicity there.}
 
{} 

\maketitle



\section{Introduction}
\label{sec:introduction}

To understand the lifecycle of the interstellar medium (ISM) we need spectrally resolved surveys that locate and characterize all the components of the ISM, including the warm ionized medium (WIM), the warm neutral medium (WNM), the atomic cold neutral medium (CNM), the diffuse molecular clouds, and the dense molecular clouds.  Carbon, which is the fourth most abundant element in the Galaxy, is very important for tracing all of these  components because its main gaseous forms: carbon ions (C$^+$), atomic carbon (C), and carbon monoxide (CO), readily emit at temperatures and densities characteristic of the ISM, and are thus not only diagnostics of the gas, but important gas coolants.  While large scale spectrally resolved surveys have been made of the Galactic diffuse atomic hydrogen gas via observation of the 21-cm  H\,{\sc i} line, and that of H$_2$ in dense UV-shielded molecular clouds with surveys of its primary tracer the CO ($J$=1$\rightarrow$0) line, much less is known on a Galactic scale about the location and characteristics of the phase where the hydrogen is molecular, but little or no CO is present.  This gas is an important component of the ISM as the carbon can readily be kept in ionized form by the interstellar radiation field (ISRF), while of order 1 to a few magnitudes of visual extinction are required for CO to become the dominant form of gaseous carbon.

CO surveys have shown that the molecular gas resides primarily in the central 1 kpc and the 4--6 kpc molecular ring \citep[e.g.][]{Dame1987, Dame2001,Clemens1988}, and is in a thin disk whose scale height is smaller than that of the atomic hydrogen gas. This localization suggests that some large-scale processes are at work in assembling the molecular clouds from the diffuse gas.  Unfortunately, tracing the details of the transition from diffuse atomic hydrogen to the dense CO molecular gas clouds has been difficult because an intermediate phase, diffuse molecular hydrogen clouds composed mainly of molecular hydrogen, are insufficiently shielded to form CO.  They are instead suffused with ionized carbon, which is impossible to observe with ground based telescopes.  Most studies of this phase have been made to date with optical and UV absorption observations of low extinction clouds, as pioneered by the {\it Copernicus} satellite, and extended by the  {\it Hubble Space Telescope} \citep[e.g.][]{Sofia2004}. However, this approach traces only lines of sight towards bright UV sources and is largely limited to local clouds ($<$1 kpc).  In contrast, C$^+$  is observed in emission throughout the Galaxy, as well as in other galaxies, in its fine-structure $^2P_{3/2}$$\rightarrow$$^2P_{1/2}$ transition, [C\,{\sc ii}], in the far-infrared at 1.9 THz (158 $\mu$m), and it is the prime probe of molecular hydrogen in photodissociation regions.  Therefore, it is an excellent tracer of the transition from diffuse atomic to dense molecular clouds.  
 
Three moderate to large-scale Galactic plane surveys of [C\,{\sc ii}] have been undertaken over the past two decades with instruments above the Earth's atmosphere on  COBE,  BICE, and  IRTS,  but with insufficient spatial and$/$or spectral resolution to isolate and characterize individual clouds along the line of sight.  The  COBE FIRAS instrument surveyed [C\,{\sc ii}] over nearly the entire Galaxy \citep[c.f.][]{Wright1991,Bennett1994}, with angular resolution of 7$\degr$ and velocity resolution $\sim$1,000 km s$^{-1}$.  COBE FIRAS found that [C\,{\sc ii}] was the strongest far-infrared spectral line in the Galaxy within the bandwidth of FIRAS (the potentially strong [O\,{\sc i}] line at 63 \micron was not observed). COBE observed that the bulk of [C\,{\sc ii}] emission arose in the inner Galaxy ($l \leq 60\degr$).  BICE, a balloon borne instrument \citep[][]{Nakagawa1998}, surveyed [C\,{\sc ii}]  with better angular resolution than COBE but only surveyed the inner region of the Galaxy, 350$\degr$ $\le ${\it l}$ \le$ 25$\degr$, {\it b} $\le$ $|3\degr|$.  BICE had an angular resolution of 15$^\prime$, and spectral resolution of $\sim$175 \kmsnospace.  Despite the much better angular resolution than COBE, the spectral resolution is insufficient to resolve individual clouds, which typically have linewidths of 1 to 10 \kmsnospace.  Nonetheless, BICE, with its 15$^\prime$ beam, was able to resolve peaks in intensity along several lines of sight. The  FILM instrument on board IRTS \citep[][]{Shibai1994,Makiuti2002} had a beam size of 8$^\prime$$\times$15$^\prime$, and a velocity resolution comparable to FIRAS ($\sim$750 km s$^{-1}$),  inadequate to resolve any velocity structure. It surveyed the Galaxy only along two 5$\degr$ wide bands forming a great circle on the sky. These three spectrally unresolved surveys reinforced the earliest model predictions that [C\,{\sc ii}]  emission is very important for the energy balance of the ISM and for tracing the ISM gas. However, without high spectral resolution they could not locate the sources along the line of sight, nor derive the physical and dynamical state of individual gas clouds through line shape and excitation analysis.

With the launch and successful operation of the {\it Herschel Space Observatory} \citep[][]{Pilbratt2010} it became possible to map   [C\,{\sc ii}] with high spectral resolution using the Heterodyne Instrument for the Far-Infrared (HIFI) instrument \citep[][]{deGraauw2010}. HIFI provides extremely high, sub-\kmsnospace, spectral resolution at 1.9 THz, and angular resolution of 12$^{\prime\prime}$, which is excellent for isolating individual cloud components that might be otherwise blended at low spectral and spatial resolution.  At this high angular resolution, and with only one pixel, {\it Herschel} is not efficient for large-scale well sampled [C\,{\sc ii}] maps, however it is well suited for sparse Galactic surveys. We have taken advantage of these capabilities of {\it Herschel} and the HIFI instrument to make the first large-scale sparse Galactic plane survey of spectrally resolved [C\,{\sc ii}] emission undertaken as part of a {\it Herschel} Open Time Key Programme, Galactic Observations of Terahertz C+ (hereafter, GOT C+).  This survey contains several hundred lines of sight of spectrally resolved [C\,{\sc ii}] emission throughout the Galactic Disk ({\it l}=0$\degr$ to 360$\degr$ and {\it b} $\le |1\degr|$). The sampling in Galactic longitude is non-uniform in order to weight as best as possible uniform angular mass, with an emphasis on the important inner $l \leq  \pm90\degr$. By analyzing a large sample of spectrally resolved components dispersed throughout the Galaxy, rather than large scale maps of a few clouds, this survey allows a statistical approach to characterizing the ISM in  [C\,{\sc ii}].   HIFI has sufficient spectral resolution to reveal the line structure of [C\,{\sc ii}] emission and separate individual ISM components on the line of sight, thus allowing us to locate them radially in the Galaxy, and describe their physical and dynamical status.  While not a map nor an image in the sense of the H\,{\sc i} and CO surveys, the GOT C+ survey does reveal the association of [C\,{\sc ii}]  emission with various cloud types and environments.  

To interpret the properties of the interstellar gas and clouds detected in [C\,{\sc ii}] we combine these spectra with H\,{\sc i} 21-cm and CO($J$=1$\rightarrow$0) isotopologue data.  This combination allows us to categorize the type of cloud we are detecting in [C\,{\sc ii}] as well as derive their gas properties and determine the relative fraction of different ISM components.   By sampling many hundreds of lines of sight and thousands of clouds, we have a statistical sample that represents the distribution of ISM  clouds.  Preliminary results from GOT C+, based on the first sixteen lines of sight taken as part of the {\it Herschel} Performance Verification Phase (PVP) and Priority Science Program (PSP), were reported by \cite{Langer2010}, \cite{Velusamy2010}, and \cite{Pineda2010} and established that  [C\,{\sc ii}] could be used to trace CO--dark H$_2$ in diffuse molecular and CO-transition clouds and measure the intensity of the FUV field in bright PDRs.  (In this paper we adopt the more accurate term CO--dark H$_2$ gas \citep[c.f.][]{Leroy2011} to refer to this ISM component, rather than ``dark gas" used by \cite{Grenier2005}, \cite{Wolfire2010}, and other authors, including ourselves in earlier papers, or the term ``dark H$_2$ gas"). \cite{Velusamy2012} also showed that along the tangent points of spiral arms, where path lengths are long, the GOT C+ data is sensitive enough to detect [C\,{\sc ii}] in the warm ionized medium (WIM) gas and characterize its properties.  

The first results using the entire GOT C+  [C\,{\sc ii}] Galactic plane survey used integrated intensities over spatial-velocity pixels (or spaxels) and performed an azimuthal average to yield the radial distribution and properties of ISM gas in the Milky Way  \citep[][hereafter Paper I]{Pineda2013}.  In the present paper we  analyze the individual clouds extracted from fitting the spectra with Gaussians. Paper I only uses the data for $b$=0$\degr$ because  only the $b$=0$\degr$ lines of sight provide a sampling throughout the entire volume of the Galaxy needed to generate a Galactic radial profile of the different ISM components. Paper I presents the first longitude--velocity maps of [C\,{\sc ii}] emission. These maps are combined with those of H\,{\sc i}, $^{12}$CO and $^{13}$CO for $b$=0$\degr$ to separate the different phases of the ISM and study their azimuthally averaged distribution and properties in the Galactic plane. In Paper I, we found that the [C\,{\sc ii}] is mainly associated with spiral arms and located between 4 to 10\,kpc. We derived the radial distribution of the cold and warm atomic, CO--dark H$_2$, and dense molecular gas. We found that the CO--dark H$_2$ extends to larger galactocentric distances (4-10\,kpc) compared to the material traced by CO and $^{13}$CO (3--8\,kpc), and that it represents about 30\% of the molecular mass of the Galaxy.

It is by now well accepted that a significant portion of the ISM gas in our Galaxy and in external galaxies is molecular hydrogen not traced by CO emission.  The presence of CO--dark H$_2$ gas in the Galaxy has been inferred from a variety of probes including dust emission \citep[c.f.][]{Reach1994,BernardPlanck2011}, $\gamma$--rays \citep[c.f.][]{Grenier2005,Abdo2010}, and C$^+$ \citep[][]{Langer2010,Langer2011,Velusamy2010,Velusamy2013,Pineda2013}.  It was also detected in external galaxies by observations of [C\,{\sc ii}]  using the Kuiper Airborne Observatory \cite[c.f.][]{Poglitsch1995,Madden1997}.  Among all these probes,  only [C\,{\sc ii}]  observations can locate the CO--dark H$_2$ gas  throughout the Galaxy because it can be observed via its $^2P_{3/2}$$\rightarrow$$^2P_{1/2}$ 1.9 THz  fine-structure transition with very high spectral resolution  using heterodyne techniques and then located using velocity--rotation curves (subject to the uncertainties of this method in the inner Galaxy for which there are two solutions -- near and far).   In the Galaxy this CO--dark H$_2$ component in massive molecular clouds is estimated  theoretically to be of order 30\%  \citep[][]{Wolfire2010} and observationally \cite[][]{Pineda2013}, but may be much larger in galaxies with lower metallicities \citep[c.f.][]{Madden2011,Madden2013PASP}.  However, the fraction of CO--dark H$_2$ will be different in less massive clouds and will also depend on its state of evolution, being essentially absent in diffuse atomic clouds and dominating the diffuse molecular clouds which are not thick enough to form significant amounts of CO. Models of the CO column densities in diffuse molecular clouds by \cite{Visser2009} show significant variations in the onset of substantial CO abundance, and thus the column density of CO--dark H$_2$ depends on physical conditions of density, temperature, and intensity of the Far-UV (FUV) radiation field.  Thus the actual fraction of CO--dark H$_2$ in the Galaxy will depend on the mass, FUV field, and evolutionary distribution of clouds.  In Paper I we found that the diffuse molecular clouds contributed most of the CO--dark H$_2$ in the Milky Way, and that the exact fraction depended on Galactic radius.

In this paper, we expand on our earlier work by analyzing the characteristics of the [C\,{\sc ii}] emission from a large ensemble of clouds extracted from the spectral data along each line of sight with Gaussian fitting routines.  We restrict our analysis to the inner Galaxy, $270\degr \leq l \leq 57\degr$  and  $|b| \leq 1\degr$ because we have observed three CO isotopologues only over  this longitude range with the ATNF Mopra 22-m telescope, and we need the rare isotopes $^{13}$CO and C$^{18}$O to better identify individual clouds where there is line blending in [C\,{\sc ii}] and $^{12}$CO.  Other CO surveys of the Galaxy do not provide us with this set of lines at the other GOT C+ lines of sight. The reduced spectra from the GOT C+ Galactic Plane survey used in this paper is available in the {\it Herschel} data archives.

Our focus in this paper is to understand the distribution of CO--dark H$_2$ among different types of clouds, including the thickness of this layer in H$_2$ column density, and the mass fraction.  In Paper I we estimated the diffuse CO--dark H$_2$ in regions with [C\,{\sc ii}] and no CO as well as where we detected [C\,{\sc ii}] and $^{12}$CO, but not $^{13}$CO.  Here, we extend the analysis of [C\,{\sc ii}] to determine the CO--dark H$_2$ in the PDRs of dense ($^{13}$CO) molecular clouds. Therefore, in this paper, we use the term CO--dark H$_2$ to refer to the gas in the FUV illuminated layer of H$_2$ in clouds, which includes the gas in clouds without detectable $^{12}$CO (diffuse molecular clouds), as well as that in the PDR envelopes of clouds with $^{12}$CO, but no $^{13}$CO  (transition molecular clouds), and those with detectable $^{13}$CO (dense molecular clouds).   
    
We begin with a summary of our observations and  then describe the extraction of the spectral features. We next present some overall statistical information on the spectral features.  We then calculate the column density and mass fraction of CO--dark H$_2$ gas by type of cloud using excitation and structural models of the clouds (discussed in an Appendix).  Finally, we present the statistical characteristics of the CO--dark H$_2$ in clouds and compare these results with models of the FUV illuminated regions of clouds. 



\section{Observations}
\label{sec:observations}

\begin{figure}
   \centering
   \includegraphics[width=8cm]{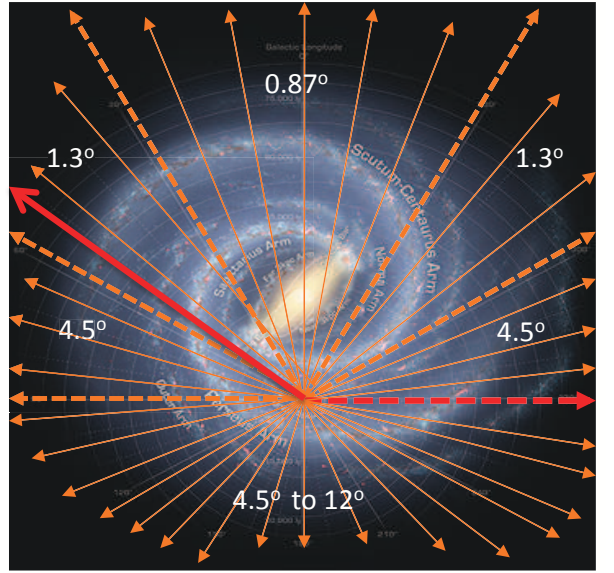}
   \caption{A schematic of the distribution of lines of sight for the GOT C+ Galactic plane survey where they are represented by solid arrows superimposed on an artistÕs impression of the Galaxy (by R. Hurt, courtesy NASA/JPL-Caltech).  Each arrow represents many lines of sight as they are too dense to display.  The longitudinal spacings are indicated on the figure and regions of different angular spacing are separated by dashed arrows.  The finest spacing is 0.87$\degr$ towards the innermost Galaxy and increases outward. The red lines mark the range {\it l}=270$\degr$ to 57$\degr$ of data analyzed in this paper. The area covered from {\it l}=90$\degr$ to 270$\degr$, which is mostly in the outer Galaxy, is only about 20$\%$ to 25$\%$ of the area of the disk and consequently a larger angular spacing is used, varying from 4.5$\degr$ at 90$\degr$ and 270$\degr$ to 12$\degr$ near 180$\degr$.}
              \label{fig:fig_1_GOTC+_LOS}
 \end{figure}

The GOT C+ programme attempts to characterize the entire Galactic disk by making a sparse sample covering 360$\degr$ in the plane. Ideally one would like to sample the disk to reproduce a uniform volume survey from the perspective of the Galactic Center.  However, from the location of the solar system in the Galaxy at a radius of 8.5 kpc, it is not possible to do so in a sparse survey.   Instead we chose observational lines of sight to provide an approximate equal angular sample of the Galaxy by mass.  In this scheme we observe at finer angular spacing towards the inner Galaxy and decrease the sampling moving outwards as shown in the schematic of the observing scheme along {\it b}=0$\degr$ in Figure~\ref{fig:fig_1_GOTC+_LOS}.  This approach produces a finer sample of the volume closest to the solar system, and the coarsest sampling at the far-side of the Galaxy.  The smallest spacing is 0.87$\degr$ towards the innermost Galaxy and increases outward.  The area covered from {\it l}=90$\degr$\ to 270$\degr$, which is outside the solar radius, is only about 20$\%$ to 25$\%$ of the area of the disk, and consequently the angular spacing in this region is larger, varying from 4.5$\degr$ to 12$\degr$ near $l =180\degr$.  

We also observed out of the plane at {\it b}=$\pm$0.5$\degr$ and $\pm$1$\degr$, alternating above and below the plane at each successive longitude.  Thus we have half as many positions surveyed at each value of {\it b} out of plane as at {\it b} = 0$\degr$, corresponding to a total of  $\sim$450 lines of sight in the disk.  Details of the observational mode of [C\,{\sc ii}] are given in Paper I along with a discussion of the data reduction.  
We also observed the $J$=1$\rightarrow$0 transitions of $^{12}$CO, $^{13}$CO, and C$^{18}$O with the ATNF Mopra 22-m Telescope  (details in Paper I), with an angular resolution of 33$^{\prime\prime}$ toward each line of sight in the inner Galaxy over a longitude range  {\it l} = $270\degr$ to $57\degr$.  We obtained H\,{\sc i} data from public sources \citep[][]{McClure2005,Stil2006}.  Table~\ref{tab:facilities} is a list of the facilities used,  angular resolution, and reference to the data source.

    \begin{table}
\caption{Observational Facilities}
\label{tab:facilities}
\begin{tabular}{l c c c c l}
\hline\hline
Tracer & Facility  & Angular Resolution &  Reference\\
  Line & & (\arcsec)  \\
\hline
[C\,{\sc ii}]& Herschel  HIFI &12 &  1,2\\
\hline
H\,{\sc i} & SGPS/ATCA &  132 &  3\\
 21-cm &VGPS/VLA& 60\  &  4\\
\hline
$^{12}$CO(1-0) & Mopra 22-m& 33 &  2\\
$^{13}$CO(1-0) &  Mopra 22-m&  35 &  2\\
C$^{18}$O(1-0) &  Mopra 22-m& 35  & 1,2\\
\hline
\end{tabular}\\

$^1$This paper;
$^2$Pineda et al. (2010 and 2013);  
$^3$McClure-Griffiths et al. (2005);
$^4$Stil et al. (2006).
\end{table}

We will be using these tracers to derive  column densities of H\,{\sc i}, CO, and [C\,{\sc ii}] in individual line of sight components. However the spatial resolution of the [C\,{\sc ii}] observations is the smallest of the tracers. Comparison of the column densities can be problematic if the H\,{\sc i} and CO emissions are inhomogenous on scale sizes much smaller than their angular resolution.  Higher resolution studies of CO show inhomogenieties down to the resolution limit \citep[c.f.][and references therein]{Falgarone2009}.  The area of the CO beam is about 7.6 times that of [C\,{\sc ii}] and the H\,{\sc i} beam is 25 to 100 times larger than that of [C\,{\sc ii}].  There is no precise way for us to estimate the filling factors in the absence of higher resolution CO and H\,{\sc i} data. However recent VLA maps of H\,{\sc i} in a small slice of the Galactic plane from the THOR (The HI/OH/Recombination line) survey do not show significant inhomogeneity on 20\arcsec scales (S. Bihr, private communication).  Furthermore, if the majority of our H\,{\sc i} data were significantly beam diluted then they would scale up to intensities and column densities that exceed those generally found in Galactic surveys \cite[c.f.][]{Dickey1990}.  

We will primarily be using $^{12}$CO to calculate column densities of the molecular gas, as this tracer is more uniform than $^{13}$CO and spatially more widespread.  Towards a few GOT C+ lines of sight there are on-the-fly cuts that show that the [C\,{\sc ii}] emission is extended over the angular size of the H\,{\sc i} and CO beams (T. Velusamy, private communication). As GOT C+ is the first large scale spectrally resolved Galactic survey of [C\,{\sc ii}] and the auxiliary data are the best available for each line of sight, it is not unreasonable to assume that the emission is uniform over the respective beams. We can compare  beam averaged intensities because we do not use them to discuss the properties of any individual clouds but instead focus on a statistical analysis of a large number of sources.
  
\begin{figure}
  \centering
   \includegraphics[width=9.0cm]{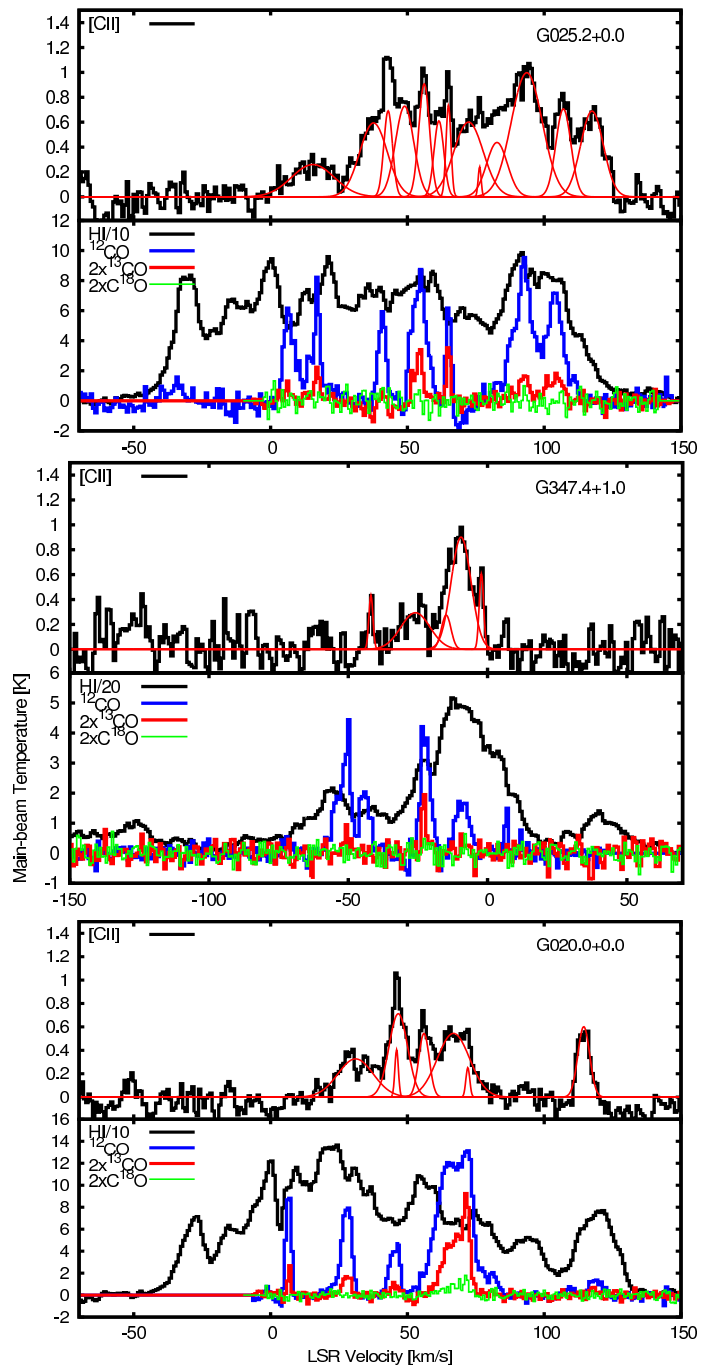}
     \caption{Spectra along three representative lines of sight from the GOT C+ inner Galaxy survey showing $T_{\rm mb}$, the main-beam temperature, versus $V_{\rm lsr}$.  The upper panel for each line of sight shows the [C\,{\sc ii}] spectrum along with the fits derived from the Gaussian decomposition (shown in red).  The lower panel for each line of sight shows the corresponding H\,{\sc i} in black, $^{12}$CO in blue, $^{13}$CO in red, and C$^{18}$O in green. The scalings for $T_{\rm mb}$ are shown in each panel.}
              \label{fig:fig_2_GOTC+_spectra}
 \end{figure}

In this paper we focus on the inner Galactic plane survey where we have a complete set of H\,{\sc i} and high angular resolution CO, consisting of 320 lines of sight covering {\it l} = $270\degr$ to $57\degr$ as indicated with red lines in Figure~\ref{fig:fig_1_GOTC+_LOS} and {\it b} $= 0\degr$, $\pm0.5\degr$, and $\pm1.0\degr$.  Note that the area used in this paper from  $270\degr$ to $57\degr$ (marked in red lines) is about 65$\%$ of the entire disk area and so gives a substantial sampling of [C\,{\sc ii}] sources in the disk especially as there are fewer detectable sources in the outer Galaxy (see Paper I). Examples of our data are shown in Figure~\ref{fig:fig_2_GOTC+_spectra}  for three lines of sight where we plot the main beam temperature of the [C\,{\sc ii}], H\,{\sc i}, and CO isotopologues, along with the Gaussian fits to the [C\,{\sc ii}] spectra (see Section 3). In many lines of sight the [C\,{\sc ii}] features are clearly blends of several components and it can be seen that a wide variety of clouds are detected in [C\,{\sc ii}] some with and some without $^{12}$CO, as well as many clouds also seen in $^{13}$CO, and even C$^{18}$O; there are also a number clouds detected in $^{12}$CO but not [C\,{\sc ii}].

\section{Spectral decomposition into [C\,{\sc ii}] clouds}
\label{sec:spectraldecomposition}

In this Section we summarize the statistical characteristics of the spectrally resolved components in the inner Galaxy.  First we describe the extraction of the cloud components from the spectral line scans, then we discuss the Galactic rotation curve model used to determine their Galactic radius, $R_{\rm gal}$.  As seen in the sample spectra in Figure~\ref{fig:fig_2_GOTC+_spectra} the distribution of [C\,{\sc ii}] in velocity is much more limited than that of atomic hydrogen (this difference is also seen in the position--velocity plots in Paper I where the [C\,{\sc ii}] spatial-velocity maps are associated with the spiral arms).  Carbon is fully ionized in diffuse atomic clouds and these are generally warm enough to excite the C$^+$ $^2P_{3/2}$ level. However, the data show that many, if not most, of the H\,{\sc i} clouds in the Galaxy have atomic hydrogen densities, $n({\rm H})$, too low to produce emission at the limits of sensitivity of our HIFI survey. Comparing the spectral (and spaxel -- see Paper I) distributions of [C\,{\sc ii}]  and $^{12}$CO(1$\rightarrow$0) we note that there are regions of  [C\,{\sc ii}]  emission without CO, and regions of CO without  [C\,{\sc ii}].  The former are likely associated with  transition molecular clouds and the latter with cold molecular clouds.  Finally, the overlap with $^{13}$CO is more restrictive, and is consistent with dense molecular clouds occupying a smaller volume of the Galaxy than the diffuse atomic and molecular clouds.


\subsection{Spectral Decomposition}

We identified the cloud components from Gaussian decompositions of [C\,{\sc ii}], $^{12}$CO, $^{13}$CO, and C$^{18}$O.  The [C\,{\sc ii}] observations were made with 0.16 \kms resolution but to improve the signal-to-noise these data were all smoothed to a velocity resolution of 0.79 \kmsnospace, as were the auxiliary data.  In Table~\ref{tab:sensitivity} we list the velocity resolution used in the analysis and typical 1-$\sigma$ noise per \kmsnospace, in some cases the rms noise for [C\,{\sc ii}] was as small as 0.06 K \kmsnospace. We also list the minimum integrated intensity (in units of K \kmsnospace) identified as an emission feature. 

The spectra towards most lines of sight are complex with many blended components and their decomposition is very labor intensive. We used an  automatic fitting routine but it had to be supplemented with individual inspection of each fit. The spectral lines were fitted with Gaussians using the IDL program XGAUSSFIT developed for use with FUSE data by \citet[][]{Lindler2001}.  We modified the program to allow up to 30 Gaussians to be fitted over the total bandwidth.  XGAUSSFIT is an interactive, least-squares, multiple Gaussian fitting routine.  It automatically fits the main component, while additional components must  be added by hand, by specifying an initial guess for the central velocity, peak antenna temperature, and Gaussian linewidth of the component.  The program then refines the fit for all of the components simultaneously.  Further details for how the routine works can be found in \citet[][]{Lindler2001}.

 \begin{table}
\caption{Observational Sensitivity}
\label{tab:sensitivity}
\begin{tabular}{l c c c c l}
\hline\hline
 Line & $\Delta$V$^a$ &  Typical 1-$\sigma$ rms &  Minimum Detection$^c$  \\
  &  [km/s] & [K km/s] & [K km/s] \\
\hline
[C\,{\sc ii}] &  0.8 & 0.08$^b$ & 0.21 \\
H\,{\sc i} 21-cm & 0.8 & 1.6 - 2 & $\sim$10 \\
$^{12}$CO(1-0) &  0.8 & 0.6 & 1.39 \\
$^{13}$CO(1-0) &   0.8 & 0.1 & 0.48 \\
C$^{18}$O(1-0) &   0.8 & 0.1 & 0.50 \\
\hline
\end{tabular}\\
(a) Velocity resolution was smoothed to this value. ( b) The lowest 1-$\sigma$ value is 0.06 K \kmsnospace. (c) The minimum intensity detected in our sample of 1804 clouds. 
\end{table}
  
For each line of sight we started by fitting Gaussians to the C$^{18}$O data as these spectra are better isolated from one another due to their relative scarcity and low opacity (see the examples in Figure~\ref{fig:fig_2_GOTC+_spectra}).  To facilitate the fitting procedure each spectral band was inspected by eye and initial guesses made for the velocity, peak, and width of the line where they could be clearly identified and then XGAUSSFIT determined the best fit.  The  central velocities and linewidths of the refined Gaussians were then used as the starting point for the fits to the $^{13}$CO spectra and then XGAUSSFIT was used to fit the remaining $^{13}$CO features, and so on, working through $^{12}$CO and finally the [C\,{\sc ii}] spectra.  At each stage additional initial inputs were provided from inspection by eye where there was no corresponding component from the previously analyzed species.  In this way we built up the spectral decomposition working from the least complex spectral tracer to the most complex ones, $^{12}$CO and [C\,{\sc ii}].  At each stage the fitting program was allowed to add additional Gaussians if necessary to fit the data. In the fitting procedure adopted here we do not consider that there may be absorption features in the spectral band, in particular for [C\,{\sc ii}] and $^{12}$CO. 

We did not apply a Gaussian fitting procedure to the H\,{\sc i} spectra because H\,{\sc i} clouds are so widely distributed in the Galaxy that the lines are a blend of many features and it is difficult, if not impossible in most cases, to extract components on a consistent basis.  Furthermore, the H\,{\sc i} 21-cm emission is mainly sensitive to column density and not local density, so there is relatively less contrast among the H\,{\sc i} spectral features.  To evaluate the intensity of the atomic component along a line of sight we use the information from the Gaussian fits to [C\,{\sc ii}] to determine the velocity range over which to calculate the H\,{\sc i} integrated intensity.  Figure~\ref{fig:fig_2_GOTC+_spectra} shows the resulting fits superimposed on the [C\,{\sc ii}] spectra along three lines of sight.  This procedure yielded over 2200 [C\,{\sc ii}] Gaussian components.  For each component the Gaussian fit gives main beam temperature, $T_{\rm mb}$, peak velocity, $V_{\rm lsr}$, FWHM linewidth,  $\Delta V$, and integrated intensity, $I=\int Tdv$.  

We then filtered these components by applying criteria on the intensity and line width.  To be a valid component we require that the integrated line intensity signal-to-noise rato be greater than 3.  We also constrain our features to have line widths in the range 1.5 to 8.0 \kmsnospace. The channel width of our smoothed data is 0.79 km s$^{-1}$ and the Gaussian fit needs a few channels, so any fits with FWHM less than $\sim$1 \kms are potentially noise spikes.  Furthermore, the full width half maximum of thermal H$_2$ is $\sim$1.3 \kms at 75\,K and of C$^+$ $\sim$0.5 \kms, so it is unlikely that the cloud components will have lines much narrower than 0.8 \kmsnospace.  To be conservative we set our lower limit for an acceptable component at 1.5 \kmsnospace.  Although the majority of the [C\,{\sc ii}] lines are well fit by a narrow Gaussian, the XGAUSSFIT procedure generates a number of very broad weak lines that represent the residual after the narrower lines are removed from the total spectrum.    About 1$\%$ of the total number of lines have features with  $\Delta V$ $>$20 \kms and about 8$\%$ have $\Delta V$ that lie in the range 10 - 20  \kmsnospace.  Almost all of these broad weak features are unlikely to be individual clouds, but could be emission from the WIM or WNM \citep[c.f.][]{Velusamy2012}, the blending of many low column density, low density CNM features, or artifacts of the fitting procedure. In this paper we take a conservative approach and exclude these  broad components from our analysis and set an upper limit of 8 \kms .  We are left with 1804 narrow [C\,{\sc ii}] lines with  1.5 \kms $\le \Delta V \le$ 8 \kms inside a Galactic radius  R$_{\rm gal} \le$9 kpc.  The 1804 [C\,{\sc ii}] components have a mean $\Delta V$ = 3.4  \kms and about $\sim$80$\%$ lie in  the range 1.5 to 6.0\,\kmsnospace.  


\subsection{Spatial location in the Milky Way}

For each Gaussian component we use the fitted $V_{\rm lsr}$  along with a Galactic rotation curve model to locate the source of the emission as a function of Galactic radius. We use a smooth model fit to tangent point velocities inside the solar radius (R$_\odot$ = 8.5 kpc), with a flat rotation curve (V$_\odot$ = 220 \kmsnospace) beyond the solar radius  \citep[c.f.][]{Johanson2009, Levine2008}.  For sources near {\it l}=0$\degr$ it is difficult to assign distances with a smooth rotation curve alone.  For these sources  we  adopt a rotation curve based on the hydrodynamical models of \cite{Pohl2008} following the procedure used by \cite{Johanson2009} and using the code given to us by C. Kerton (private communication).  \cite{Pohl2008} use a gas-flow model derived from smoothed particle hydrodynamics (SPH) simulations in gravitational potentials based on the near infrared (NIR) luminosity distribution of the bulge and disk \citep{Bissantz2003}.  This approach  provides a more accurate kinematics of the clouds in the inner Galaxy especially in the range $-6\degr \leq l \leq +6\degr$. This model allows us to use all the observed velocity components including those which are not consistent with realistic solutions from a simple rotation curve.  Over most of the range of {\it l} within the solar circle discussed here, for a given $V_{\rm lsr}$  and {\it l},  the rotation curves yield near and far distances and a unique value for the Galactic radial distance.  We do not attempt to resolve the near--far distance ambiguity as we are interested only in the radial distance from the Galactic center.  

\begin{figure}[!htb]
     \centering
     \includegraphics[width=7.0cm]{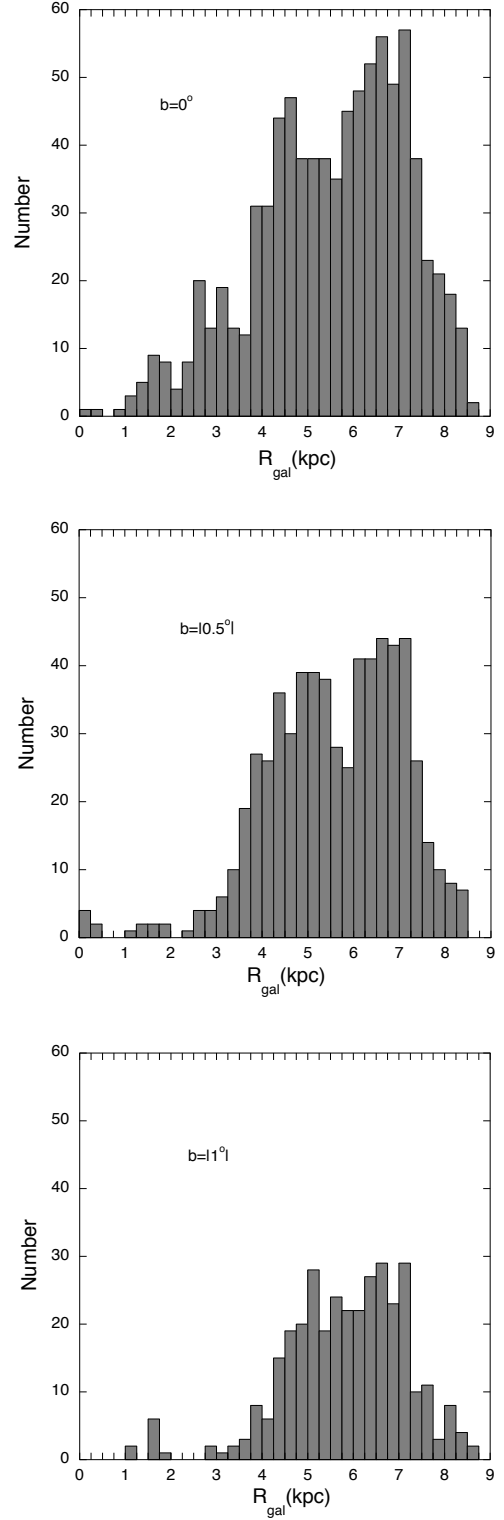}
     \caption{The number distribution of [C\,{\sc ii}] sources as a function of Galactic radius for $b = 0\degr$, $\pm 0.5\degr$ combined, and $\pm 1\degr$ combined.  With increasing $b$ there are fewer sources and their distribution skews closer to the solar radius because the line of sight at some distance away from the solar system is above the plane where there a fewer sources; see text for details.}
      \label{fig:fig_3_CII_sources_R_b}
 \end{figure}

In Figure~\ref{fig:fig_3_CII_sources_R_b} we plot the radial distribution of [C\,{\sc ii}] sources for $|b|=0\degr$, 0.5$\degr$, and 1$\degr$.  The majority of the sources are located between 4 and 7.5 kpc and the median values are 5.8,  5.7, and 6.0 kpc for $|b|=0\degr$, 0.5$\degr$, and 1$\degr$, respectively.  For all sources combined the median radius is 5.8 kpc. The distributions are affected by two factors, the angular sampling and the value of  $b$.  As mentioned in Section 2, for sources with $b\neq 0\degr$ the line of sight samples the volume near the solar system better than further away, such that more [C\,{\sc ii}] and CO sources will be intercepted closer to the solar system.  This unavoidable factor will skew the radial distribution to Galactic radii closer to the solar system, and is responsible for some of the shift in the distribution. The distributions for $|b| = 0.5\degr$ and $1\degr$ are slightly narrower in Galactic radius than that for $b=0\degr$ and there is a slight skewing of the data for $|b|=1\degr$ with relatively fewer sources inwards of 4 kpc. This shift for $b\ne$ 0$\degr$ is due to the line of sight being at a distance away from the plane greater than the thickness of the Galactic disk at some distance from the solar system  and thus not uniformly sampling clouds across the Galaxy.  The thickness of the [C\,{\sc ii}]  gas in the Milky Way is estimated to be of order 150 pc (Paper I) so that near $l = 0\degr$ sources inward of R$_{\rm gal}$=4 kpc at $b = \pm 1\degr$ are $\sim$80 pc above or below the plane. This undersampling at $|b| \ge$0.5$\degr$  is reflected in the decrease in the distribution of clouds seen in  Figure~\ref{fig:fig_3_CII_sources_R_b}.

\subsection{Distribution of clouds by gas tracers}

We have catalogued all the [C\,{\sc ii}] clouds and labeled them according to the presence or absence of $^{12}$CO and $^{13}$CO, which depends on the sensitivity of our survey to these isotopologues. We require that all CO spectra have an integrated intensity with a signal-to-noise greater than 3.  The typical rms noise  and minimum detected intensity for $^{12}$CO, $^{13}$CO, and C$^{18}$O in our data base of 1804 clouds are given in 
Table~\ref{tab:sensitivity}.   The corresponding threshold column densities needed to detect these CO isotopologues was estimated using  the approximate formulas in \cite{Carpenter2000} assuming optically thin thermalized lines (these were also checked against the RADEX code \cite[][]{vanderTak2007} using typical densities (n(H$_2$) $\sim$300 to 1000 cm$^{-3}$) and temperatures (10 to 40K)).  The resulting minimum column densities are  $N$($^{12}$CO) $\sim$1-2$\times$10$^{15}$ cm$^{-2}$, and for $N$($^{13}$CO) and $N$(C$^{18}$O)  $\sim$4-8$\times$10$^{14}$ cm$^{-2}$. The majority of the CO isotopologue lines are much stronger than these threshold values.  In Appendix A.4 we discuss some properties of the relationship among the $^{12}$CO and $^{13}$CO isotopologue intensities.  

Clouds with no detectable CO are labeled diffuse molecular  (H$_2$) clouds (as was shown in  \cite{Langer2010} and \cite{Velusamy2010} very few  [C\,{\sc ii}] sources are diffuse atomic H\,{\sc i} clouds), while those with $^{12}$CO, but no $^{13}$CO are transition molecular clouds, and the clouds with $^{13}$CO are referred to as dense molecular clouds because they have higher column and, likely, volume densities (those with C$^{18}$O may be even more highly evolved with denser interiors and cores).  In Table~\ref{tab:clouds} we summarize the number of lines of sight and number of clouds by their classification.  For all the narrow line [C\,{\sc ii}]  and CO isotopologue data  we created a data base of longitude, latitude, $V_{\rm lsr}$, FWHM, integrated intensities, and Galactic radius, $R_{\rm gal}$.   As discussed above the  21-cm H\,{\sc i} spectra are so strongly blended that it is not, in most cases, possible to extract them individually using the Gaussian fitting technique adopted for [C\,{\sc ii}]  and CO.  Later on we will need to know the column density of atomic hydrogen towards each [C\,{\sc ii}] source.  For the atomic hydrogen tracer we calculate the H\,{\sc i} intensity in each cloud by integrating within the velocity widths defined by the [C\,{\sc ii}] and/or CO lines. 


\begin{table}
\caption{Number of  [C\,{\sc ii}] clouds by type and Galactic latitude}
\label{tab:clouds}
\begin{tabular}{c c c c c c}
\hline
\hline
  &  & $b$= 0$\degr$ & $\pm$0.5$\degr$ & $\pm$1.0$\degr$ & Total \\
  & Lines of Sight  & 114 & 113 &  93 & 320 \\
 \hline
 \hline
 Tracer & Cloud Classification &  &  &  & \\
  \hline
No CO & Diffuse Molecular  & 194 & 218 & 145 & 557 \\
$^{12}$CO & Transition  Molecular  & 214 & 187 & 112  &  513 \\
$^{13}$CO & Dense  Molecular & 313 & 144 & 65 & 522  \\
C$^{18}$O & Dense Cores   & 116 & 74 & 22 &  212 \\
\hline
Total &  [C\,{\sc ii}]  Clouds & 837 & 623 & 344 & 1804 \\
\hline
\hline
$^{12}$CO  & No  [C\,{\sc ii}]   & 280 & 241 & 193  & 714 \\
\hline\hline
\end{tabular}
\end{table}

\begin{figure*}[t]
     \centering
     \includegraphics[scale=1.0]{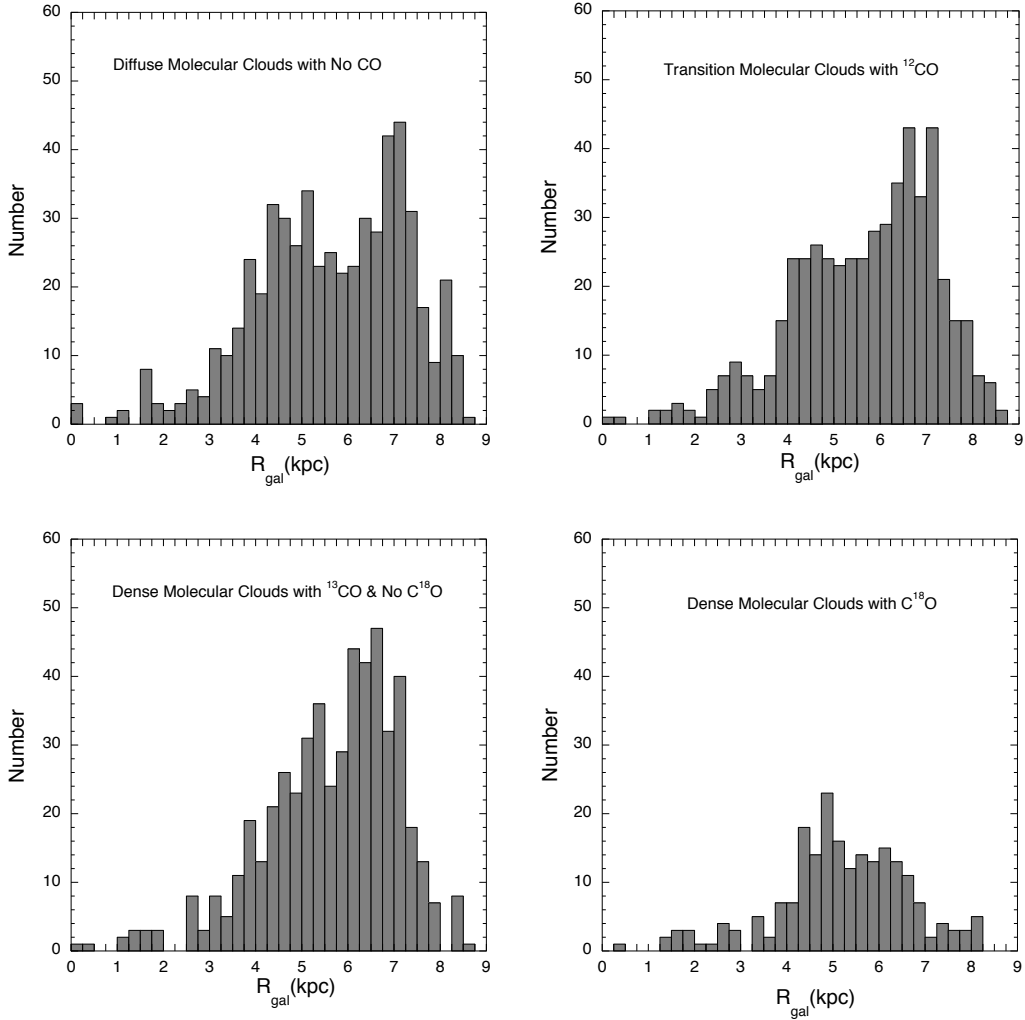}
    \caption{The distribution of [C\,{\sc ii}] sources as a function of Galactic radius for clouds as characterized by the presence or absence of CO isotopologues. All values of Galactic latitude, {\it b}, are included in this figure.}
        \label{fig:fig_4_CII_R_CO_type}
\end{figure*}

In Figure~\ref{fig:fig_4_CII_R_CO_type} we plot the number distribution of [C\,{\sc ii}] clouds by category labeled diffuse molecular, transition molecular, and dense molecular cloud,  as a function of Galactic radius; in the case of the dense molecular clouds we further separate them by the presence or absence of C$^{18}$O.  The distributions of the different cloud categories are generally similar.  They are mostly confined to R = 4 to 7.5 kpc, although the diffuse clouds (no CO) are more extended beyond 7.5 kpc, while the densest clouds, those containing C$^{18}$O are more confined to 4.5 to 6.5 kpc. As discussed for Figure~\ref{fig:fig_3_CII_sources_R_b}, the line of sight sampling can skew the distribution to enhance the number of sources closer to the solar system and therefore at larger Galactic radii within the solar circle.
 
\begin{figure}
     \centering
        \includegraphics[width=7cm]{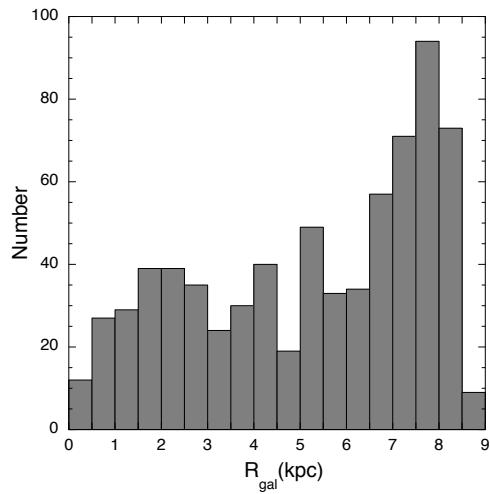}
     \caption{The distribution of CO sources without detectable  [C\,{\sc ii}] as a function of Galactic radius; all values of Galactic latitude, {\it b}, are included in this figure.}
         \label{fig:fig_5_CO_R_no_CII}
 \end{figure}

In addition to these sources, we have 714 clouds detected  in $^{12}$CO  in the inner Galaxy without any corresponding [C\,{\sc ii}] (at the 4-$\sigma$ level).  Their distribution by latitude  is given in Table~\ref{tab:clouds} and  displayed by Galactic radius in Figure~\ref{fig:fig_5_CO_R_no_CII}.  In contrast to the CO sources with  [C\,{\sc ii}], the distribution of CO sources without [C\,{\sc ii}] is skewed towards larger Galactic radii 7 to 9 kpc.  Two factors contributing to this distribution, similar to that seen in [C\,{\sc ii}], is that for $b= \pm 0.5\degr$ and $\pm 1\degr$, clouds above the plane and inside of $R_{\rm gal}$ about 4 kpc would be missed along these lines of sight and the volume near the solar system is better sampled.

 \section{CO--dark H$_2$ in clouds}
 \label{sec:CO--darkH2}

In \cite{Langer2010} and \cite{Velusamy2010} we showed, based on a limited cloud sample, that the observed emission of [C\,{\sc ii}] could not be explained by diffuse atomic clouds, and that it requires a significant fraction of  [C\,{\sc ii}] coming from FUV illuminated molecular gas -- the CO--dark H$_2$.  In \cite{Velusamy2013}  we confirmed this result based on preliminary extraction of a larger data set from the complete survey. Using the complete GOT C+ data set (Paper I) we calculated the azimuthally averaged radial distribution of ISM gas components including the CO--dark H$_2$.  Here we expand on these earlier studies and analyze the column densities of different gas components, as traced by H\,{\sc i,} [C\,{\sc ii}], and CO, on a cloud--by--cloud basis using our sample of 1804  [C\,{\sc ii}] clouds.  First we re-examine the relationship of [C\,{\sc ii}] and H\,{\sc i} for these clouds showing that the observed [C\,{\sc ii}] cannot arise solely from diffuse atomic gas for the vast majority of the clouds.

\begin{figure}
  \centering
 \includegraphics[width=8cm]{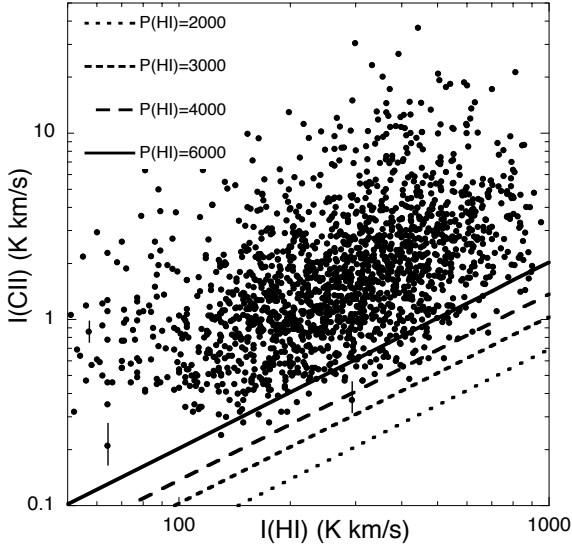}
      \caption{Scatter plot of $I$([C\,{\sc ii}]) versus $I$(H\,{\sc i}) for all [C\,{\sc ii}] sources.  The straight lines are predicted $I$([C\,{\sc ii}]) for low density atomic H\,{\sc i} diffuse clouds as a function of $I$(H\,{\sc i}) (which is proportional to the H\,{\sc i} column density) for four different characteristic pressures from $P = 2000$\,K cm$^{-3}$ to 6000 K cm$^{-3}$. This pressure range covers predicted values for the diffuse clouds from $R_{\rm gal}$ $\sim$10 kpc to  $\sim$4 kpc, with 3000 K cm$^{-3}$ characteristic of the solar radius.  It is clear that the vast majority of the GOT C+ [C\,{\sc ii}] sources require an additional source of [C\,{\sc ii}] emission other than the diffuse atomic gas. The 1-$\sigma$ noise for $I$([C\,{\sc ii}])  is indicated on three representative low intensity points; those for $I$(H\,{\sc i}) are too small to be visible on this figure and have been omitted.  It can be seen that on the scale of this figure that the S$/$N is high for most of points in this scatter plot.}
              \label{fig:fig_6_CII_vs_HI_P(HI)}
    \end{figure}

In Figure~\ref{fig:fig_6_CII_vs_HI_P(HI)} we plot on a log-log scale the integrated intensity $I$([C\,{\sc ii}])=$\int T($[C\,{\sc ii}]$dv$, versus that for  H\,{\sc i}, $I$(H\,{\sc i})=$\int T($H\,{\sc i}$)dv$ for all sources.  We note several characteristics of this scatter plot.  First, $I$([C\,{\sc ii}]) is detected even for very small values of  $I$(H\,{\sc i})$\le$100 K \kmsnospace, corresponding to a low column density of diffuse atomic hydrogen $N$(H\,{\sc i})$\le$1.8$\times$10$^{20}$ cm$^{-2}$. Second, the largest values of $I$(H\,{\sc i}) yield a column density, $N$(H\,{\sc i}) $\sim$2$\times$10$^{21}$ cm$^{-2}$, or about A$_{\rm V}$=2 mag. The column density for typical H\,{\sc i} diffuse clouds is less than a few $\times$10$^{20}$ cm$^{-2}$ and for the atomic envelopes of molecular clouds it ranges from $\sim$1 to 7$\times$10$^{20}$ cm$^{-2}$ (see Figure 7 in \citealt{Wolfire2010}).  Thus this emission probably arises from several such clouds and/or the H\,{\sc i} envelopes around dense molecular clouds in proximity along the line of sight.  Third,  for any given value of $I$(H\,{\sc i}) there is a large scatter in $I$([C\,{\sc ii}]) indicating that the [C\,{\sc ii}] is probing a wide range of cloud conditions and column densities.  
 
We cannot rule out that some of the scatter could be due to the different beam sizes used to observe [C\,{\sc ii}] and H\,{\sc i}, because, as the beam sizes are very different, there may be detections where we underestimate the H\,{\sc i} intensity if the emission is very nonuniform. However, we do not believe that beam dilution is a large factor in Figure~\ref{fig:fig_6_CII_vs_HI_P(HI)} for two reasons. First, as discussed in Section 2, VLA 21-cm observations from the THOR survey do not show significant variations on a 20$^{\prime\prime}$ scale. Second, the highest H\,{\sc i} intensities in our map $\sim$10$^3$ K \kms (corresponding to a column density $\sim$2$\times$10$^{21}$ cm$^{-2}$; see Appendix A.1) and,  if the filling factor correction were large, it would shift many points outside the range of intensities found in Galactic H\,{\sc i} surveys \cite[c.f.][]{Dickey1990}. Furthermore, in a few locations there are on-the-fly [C\,{\sc ii}] maps that show that the [C\,{\sc ii}] emission is extended over the angular size of the H\,{\sc i} beam (T. Velusamy, private communication). Finally, our analysis is statistical in nature and some $N$(H\,{\sc i}) may be underestimated while others are overestimated, but on average should represent the characteristics of the H\,{\sc i} to [C\,{\sc ii}] ratio. Therefore we assume we can use the beam average values of $I$(H\,{\sc i}) and $I$([C\,{\sc ii}]) for this comparison.

We have modeled the expected $I$([C\,{\sc ii}]) as a function of $I$(H\,{\sc i}), similar to what was done in \cite{Langer2010} and Paper I.   In those papers we used $I$(H\,{\sc i}) to calculate $N$(H\,{\sc i}), and then we multiplied it by the fractional abundance of C$^+$ with respect to atomic hydrogen, $x_{\rm H}$(C$^+$), to determine the column density of C$^+$ in the H\,{\sc i} gas, $N_{\rm HI}$(C$^+$).  We used a value of $x_{\rm H}$(C$^+$) = 2.2$\times$10$^{-4}$ characteristic of the metallicity at of 5.8 kpc, the median Galactic radius for all those sources (see Sections 3.2 and 4.4 for discussions of the radial distribution and metallicity, respectively). Here we calculate $I$([C\,{\sc ii}])  from the excitation equation (see Appendix A)  for a given set of thermal pressures, corresponding to different densities and temperatures, and the same representative metallicity $x_{\rm H}$(C$^+$)=2.2$\times$10$^{-4}$.   In Figure~\ref{fig:fig_6_CII_vs_HI_P(HI)} we plot  $I$([C\,{\sc ii}]) for a family of ($n({\rm H})$,$T_{\rm kin}$) which are labeled with the corresponding pressure, $P=nT_{\rm kin}$ (in units of K\,cm$^{-3}$).  We fixed the kinetic temperature at a typical value for H\,{\sc i} clouds, $T_{\rm kin}$=100\,K.  For an appropriate range of thermal pressure we were guided by the work of \cite{Wolfire2003}.  They derived the thermal pressure of low density, low column density atomic clouds based on two-phase thermal pressure equilibrium model of CNM in equilibrium with the WNM, taking into account the variation in metallicity and visible-UV radiation across the Galaxy.  As a function of Galactic radius, $R_{\rm gal}$ in kpc, the pressure is fit by

\begin{equation}
P_{\rm HI}(R_{\rm gal})=1.4\times 10^4 e^{-R_{\rm gal}/5.5}\, {\rm K\,cm^{-3}}.
\end{equation}

\noindent  For R$_{\rm  gal}$ $\sim$ 4.5 kpc to 8.5 kpc, where most sources are located, the pressure lies in the range $\sim$6200 to 3000 K cm$^{-3}$.  We display the predicted $I$([C\,{\sc ii}]) in Figure~\ref{fig:fig_6_CII_vs_HI_P(HI)}, for this wide range of typical atomic cloud thermal pressures from $P\sim$3000 K cm$^{-3}$ near the solar radius, R$_{\odot}$, to $\sim$6000 K cm$^{-3}$ at the inner radius of the molecular ring.  For the GOT C+ sample the median radius of the distribution in Figure~\ref{fig:fig_3_CII_sources_R_b} (top panel) is about 5.8 kpc, corresponding to a thermal pressure of $\sim$4900 K cm$^{-3}$ for H\,{\sc i} clouds in the \cite{Wolfire2003} model. The observed $I$([C\,{\sc ii}]) are in general much greater than the values predicted for diffuse atomic clouds.   We assume we can use the beam average values of $I$(H\,{\sc i}) and $I$([C\,{\sc ii}]) for this comparison, however as the beam sizes are very different there may be detections where we underestimate the H\,{\sc i} intensity on the scale of the [C\,{\sc ii}] beam if the emission is very nonuniform. Furthermore, to change our results significantly a large fraction of the samples would have to be significantly beam diluted in H\,{\sc i}. We conclude that H\,{\sc i} is not a significant collision partner for exciting C$^+$.  This conclusion is supported by the fact that [C\,{\sc ii}] is not detected over a large ($\sim$70$\%$) fraction of the H\,{\sc i} emission seen in the wide velocity range over the entire band in all lines of sight, as illustrated in the three examples in Figure~\ref{fig:fig_2_GOTC+_spectra}. In addition, a spaxel by spaxel analysis of the correlation of H\,{\sc i}, [C\,{\sc ii}] , and CO shows that the GOT C+ survey does not detect much [C\,{\sc ii}]  from low density H\,{\sc i} clouds (T. Velusamy, private communication). Thus, the majority of the observed [C\,{\sc ii}] in the narrow spectral features comes from FUV illuminated molecular gas, that is the CO--dark H$_2$ gas layer, in clouds.  The model for [C\,{\sc ii}] emission coming from both H\,{\sc i} and H$_2$ gas along the line of sight is discussed in more detail below.

Thus, to determine the properties of the [C\,{\sc ii}]  clouds we need to calculate the column densities of the three main cloud layers as traced by: H\,{\sc i}, [C\,{\sc ii}], and CO corresponding to each spectral feature.  We assume that most of the H\,{\sc i} column density comes from the outermost gas layer in the cloud, but some emission might come from nearby gas along the line of sight within the same velocity range, but not physically associated with the cloud.  Such purely atomic H\,{\sc i} clouds are likely to be at lower densities than that in the outer layers of denser clouds and their contribution to the total column density can probably be neglected.  Below we describe in more detail how we determine the material in each layer. Because we do not map the clouds the results are valid for only one line of sight with the given resolution in each cloud and may not be representative of the cloud as a whole. However, GOT C+ detects so many spectral components that it should represent a statistical sampling of ISM clouds.  The assumption adopted here is that the line of sight results represent the properties of the cloud. 


 \subsection{CO--dark H$_2$ Model Calculation}
 
\begin{figure}
 \includegraphics[width=9cm]{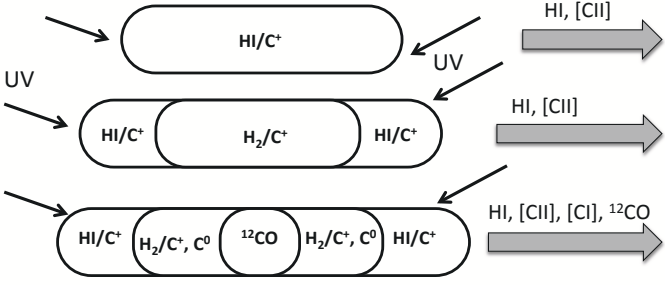}
      \caption{Schematic of the [C\,{\sc ii}] emission from different types of clouds along the line of sight.}
              \label{fig:fig_7_cloudlayers}
    \end{figure}

Figure~\ref{fig:fig_7_cloudlayers} shows a schematic of what is observed along the line of sight in $I$([C\,{\sc ii}]).  This cartoon illustrates three of the line of sight environments sampled by our survey: (top) diffuse atomic hydrogen clouds containing C$^+$ that emit in H\,{\sc i} 21-cm and [C\,{\sc ii}] 158 $\mu$m$\space$ radiation; (middle) diffuse molecular clouds with atomic and molecular hydrogen layers containing C$^+$ in which there is  [C\,{\sc ii}] emission from both regions; and, (bottom) transition molecular clouds containing atomic and molecular hydrogen, with unshielded H$_2$ (C$^+$), and shielded H$_2$ (containing $^{12}$CO) layers. The transition molecular clouds are traced by 21-cm H\,{\sc i}, 158 $\mu$m [C\,{\sc ii}], and 2.6 mm CO(1$\rightarrow$0) emission (there is a small neutral carbon layer, or mixed layer containing CO and C$^+$, which does not contain much gas and we omit for simplicity and lack of [C\,{\sc i}] data). There are also dense molecular clouds (not shown) which contain, in addition to emission from the $^{12}$CO layer in the bottom panel, a deeper layer of $^{13}$CO emission and sometimes  at larger depths C$^{18}$O cores.

This schematic and our subsequent analysis ignores the contribution to [C\,{\sc ii}] from the warm ionized medium (WIM) for several reasons.  First, in Paper 1 we estimated that at {\it b}=0$^{\rm o}$ the WIM contributes on average about 4$\%$ to the total intensity of the [C\,{\sc ii}] emission.  Second, while the relative contribution of the ionized gas can be much larger towards H\,{\sc ii} regions or very bright PDRs,  the GOT C+ survey intersects very few, if any, of these regions.  Third, the filling factor for the WIM and WNM is larger than that for the CNM and thus the WIM and WNM features arise from a longer path length and thus have a broader line width.  In \cite{Velusamy2012} we were able to detect  directly the [C\,{\sc ii}] emission from the warm ionized medium, but only along the tangent points in the plane where  a relatively small velocity dispersion corresponds to a very long path length, and only at the edge of a spiral arm where the densities are a factor of 2 to 4 higher than in the general interstellar WIM.  In other words the intensity per unit velocity is relatively large along the tangent points, whereas towards other lines of sight the emission per unit velocity is much smaller and the WIM (and WNM) does not contribute much [C\,{\sc ii}] within the line width characterizing each component in our sample, which are restricted to $<$8\kmsnospace.  (The WIM is also more important above the plane where its filling factor is much larger, but our narrow line [C\,{\sc ii}] components are restricted to $\pm$ a few hundred parsecs within the disk.)  Thus, as noted in Section 3.1, our Gaussian decomposition of  [C\,{\sc ii}] spectra resulted in residual emission that was very broad but weak, and so removes some of the emission that might arise from the WIM as we only consider the narrow line [C\,{\sc ii}] components.

We estimated the fraction of CO--dark H$_2$ gas in our GOT C+ data base using the layered cloud models in Figure~\ref{fig:fig_7_cloudlayers}. We first determined how much [C\,{\sc ii}]  emission comes from the H\,{\sc i} gas,  $I_{\rm {HI}}$([C\,{\sc ii}]), and then subtract that from the observed total $I_{\rm tot}$([C\,{\sc ii}])  emission.  From $I_{\rm{H_2}}$([C\,{\sc ii}]) we can calculate $N_{\rm{H_2}}({\rm C^+})$ and then $N_{\rm {C^+}}$(H$_2$)=$N_{\rm{H_2}}({\rm C^+})/x_{\rm H_{2}}$(C$^+$), where $x_{\rm H_{2}}$(C$^+$) is the C$^+$ fractional abundance with respect to H$_2$, provided we have an estimate of the densities, temperatures, and fractional carbon abundance in these layers.  We determined the contribution of the  atomic hydrogen on the line of sight  to the total intensity, $I_{\rm tot}$([C\,{\sc ii}]) by calculating $N$(H\,{\sc i}) from $I$(H\,{\sc i}) as described above and in Appendix A.  Then we convert it to the corresponding $N_{\rm HI}$(C$^+$) using the appropriate metallicity for the location in the Galaxy and then determine $I_{\rm HI}$([C\,{\sc ii}])  given the radiative transfer equation in Appendix A, assuming appropriate temperature and density. We set the kinetic temperature to 100\,K, characteristic of that measured for H\,{\sc i} clouds -- the result is not particularly temperature sensitive over the range measured for diffuse clouds.  We choose a density from the thermal pressure in the atomic hydrogen gas, $n$(H\,{\sc i})=$P/T_{\rm kin}$, using the pressure gradient in the Galaxy $P_{\rm HI}$($R_{\rm gal}$) in Equation (1) above (see also Paper I) derived by  \cite{Wolfire2003}.  

The following list summarizes the steps in our model approach:

 \begin{enumerate}
      \item Use the velocity information and rotation curve to determine the Galactic radius, $R_{\rm gal}$, for each [C\,{\sc ii}] and CO source; 
      \item Determine $N$(H\,{\sc i}) from $I$(H\,{\sc i}) over the velocity range of the [C\,{\sc ii}] line;
     \item  Derive $N_{\rm HI}$(C$^+$) = $x_{\rm HI}$(C$^+$)$N$(H\,{\sc i}) using the appropriate Galactic radial value of $x_{\rm HI}$(C$^+$);
     \item  Determine the corresponding [C\,{\sc ii}]  emission intensity, $I_{\rm {HI}}$([C\,{\sc ii}]), from the H\,{\sc i} layer assuming $T_{\rm kin}$=100\,K and $n({\rm H})=P_{\rm HI}(R_{\rm gal})T^{-1}$; 
     \item From the total [C\,{\sc ii}] intensity, $I_{\rm total}$([C\,{\sc ii}]), subtract the emission coming from the atomic H, yielding the [C\,{\sc ii}] emission from the molecular gas,  $I_{\rm H_{2}}$([C\,{\sc ii}]) =  $I_{\rm total}$([C\,{\sc ii}])  - $I_{\rm {HI}}$([C\,{\sc ii}]);
    \item Use $I_{\rm H_{2}}$([C\,{\sc ii}]) to calculate the column density $N_{\rm {H_2}}({\rm C^+})$ using the radiative transfer equation in Appendix A, assuming $T_{\rm kin}$(H$_2$) and $n$(H$_2$) from a pressure model of the CO--dark H$_2$ layer (see below);
   \item Derive the column density of H$_2$ traced by [C\,{\sc ii}],  $N{\rm_{[CII]}}$(H$_2$) = $N{\rm_{H_2}}$(C$^+$)/$x_{{\rm H_2}}$(C$^+$);
   \item If $^{12}$CO is present, then calculate the H$_2$ column density in the CO interior using the CO--to--H$_2$ conversion factor, $N_{{\rm CO}}$(H$_2$)=$X_{\rm {CO}}I(^{12}$CO).
   \end{enumerate}
     
\noindent The details of the relationships and formulas used to calculate the column densities are given in Appendix A.  After applying these procedure to all of the [C\,{\sc ii}]  spectra we have the column densities for each source H\,{\sc i}, $N$(H\,{\sc i}), CO--dark H$_2$, $N$(CO--dark H$_2$) = $N_{\rm [CII]} ({\rm H_2})$, and CO traced H$_2$,  $N$(CO--traced H$_2$)=$N_{{\rm CO}}$(H$_2$). The total column density of H$_2$ gas, $N_{\rm tot}$(H$_2$)=$N$(CO--dark H$_2$) + $N$(CO--traced H$_2$), combines the contribution of the CO--dark H$_2$ and CO--traced H$_2$ layers.

 
\subsection{Density and temperature in the clouds} 

To calculate the column density $N_{\rm H_{2}}$(C$^+$) from $I_{\rm H_{2}}$([C\,{\sc ii}]) we need to know the temperature and density of the CO--dark H$_2$ gas (see Appendix A for radiative transfer equations).  Unfortunately, C$^+$ has only one fine-structure transition so there is no self-consistent way to determine $N_{H_{2}}$(C$^+$),  $n(H_2)$, and $T_{\rm kin}$ simultaneously.  Without information on the density and temperature of the H$_2$ gas we can only determine a combination containing $N({\rm C}^+)$, $n({\rm H}_2)$, and $T_{\rm kin}$, as can be seen by rewriting Equation A.7 in the Appendix as, 

\begin{equation}
I_{H_{2}}([{\rm CII}]) = 6.9\times10^{-16} N_{H_{2}}({\rm C^+})\frac{n(H_2)}{n_{\rm cr}(H_2)}e^{-{\Delta}E/kT}\, (\rm K\, km\, s^{-1})
\end{equation}

\noindent where $\Delta E/k$= 91.25 K.  We have found that it is possible to combine the dependence on density and temperature and express the column density approximately in terms of just the thermal pressure, $P/k=n(H_2)T(H_2)$ in units of (K cm$^{-3}$) times a term that is a weak function of T$_{\rm kin}$ over the temperature range of interest.  Then by specifying the thermal pressure as a function of Galactic radius or with an independent measure of thermal pressure we can estimate the column density of CO-dark H$_2$. If we assume a constant thermal pressure in the [C\,{\sc ii}]  emitting layer we can replace the density by  $n(H_2)=P(H_2)/T$ where the pressure is in units of K cm$^{-3}$.  As the critical density is a very weak function of kinetic temperature over the temperature range of interest here \citep{Goldsmith2012} it can be set to a constant and we can rewrite Equation 2 as, 

\begin{equation}
I_{H_{2}}([{\rm CII}]) = 6.9\times10^{-16} \frac{P(H_2)N_{H_{2}}({\rm C^+})}{n_{\rm cr}(H_2)}\frac{e^{-{\Delta}E/kT}}{T}.
\end{equation}

\noindent (The term $PT^{-1}exp({-{\Delta}E/kT})$ is related to the collision rate from the lower to upper state.)

\begin{figure}
 \includegraphics[width=8cm]{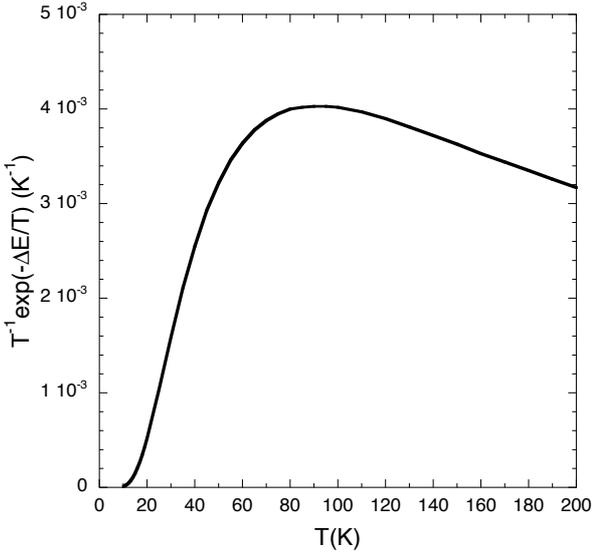}
      \caption{Dependence of $T^{-1}exp(-91.25/T)$, which is proportional to the relative excitation rate to the $^2P_{3/2}$ level of C$^+$, as a function of temperature.  This term in the excitation rate varies by only $\pm$15$\%$ about its mean value of 3.5$\times$10$^{-3}$ over the range  45\,K$\le$T$_{\rm kin}$$\le$200\,K, and by only $\pm$20$\%$ from 40 K to 200K.}
              \label{fig:fig_8_CII_TvsP}
    \end{figure}

In Figure~\ref{fig:fig_8_CII_TvsP} we plot the expression $T^{-1}exp(-91.25/T)$  as a function of $T$. 
It can be seen that it varies by less than about $\pm15\%$ about its mean value of 3.5$\times$10$^{-3}$ 
over a broad range of temperatures 45\,K$\le$$T_{\rm kin}$$\le$200\,K and is good to about $\pm 20\%$ down to 40\,K).  
Thus if we have an estimate of the thermal pressure of the H$_2$ layer, the
choice of $T_{\rm kin}$ is not critical over this temperature range. At less than 40\,K the excitation rate drops sharply and the choice of
$T_{\rm kin}$ is important.  We believe it is unlikely much of the [C\,{\sc ii}] emission detected by GOT C+ comes from gas with  T$_{\rm kin}<$35\,K,  although the survey is sensitive to emission down to 20 K depending on pressure and column density.  For example, the GOT C+ survey can detect [C\,{\sc ii}] in gas down to 45 K for the diffuse molecular clouds (no CO) with column densities at the lower end of their range, $N({\rm C^+})\sim10^{17}$ cm$^{-2}$, for typical diffuse gas pressures.  For the  median column density of these clouds, $N({\rm C^+})\sim3\times10^{17}$ cm$^{-2}$ GOT C+ is sensitive to gas at temperatures down to $\sim$25 K.  However, for the dense molecular clouds with high pressure PDRs the GOT C+ survey can detect [C\,{\sc ii}] emission down to $\sim$15 K.   Thus for warm C$^+$ layers, $>$40 K, the column density is mainly a function of the thermal pressure in the gas and the column density can be written approximately in the following form using the mean value above for $T^{-1}exp(-91.25/T)$, 

\begin{equation}
 N_{H_{2}}({\rm C^+})\simeq 4.2\times10^{17} \frac{n_{\rm cr}(H_2)}{P(H_2)} I_{H_{2}}([{\rm CII}]);\,40K<T_{kin}<200K
\end{equation}

\noindent where $N_{H_{2}}({\rm C^+})$ is in cm$^{-2}$ and this solution is good to about $\pm$20$\%$ over the indicated temperature range. Furthermore, it is reasonable to assume that the thermal pressure is roughly constant across the CO--dark H$_2$ layer, because, as C$^+$ is the primary coolant in the layer, if the density increases the temperature decreases.  Thus if we can characterize the thermal pressure as a function of cloud type and Galactic environment, then we can solve for the column density $N_{\rm H_{2}}$(C$^+$).  Note that Equations (3) and (4) were derived to demonstrate that use of a constant pressure in the C$^+$ layer is reasonable.  However, for all calculations of [C\,{\sc ii}] intensity and C$^+$ column density,  we use the full radiative transfer equations in Appendix A. Then $N$(H$_2$) traced by [C\,{\sc ii}] is just,

\begin{equation}
N_{\rm [CII]} ({\rm H_2})=  N_{\rm{H_2}}({\rm C^+})/ x_{{\rm H_2}}(\rm C^+)
\end{equation}

\noindent where $x_{\rm H_2}(\rm C^+) = 2x_H(\rm C^+)$, as H$_2$ contains two atoms of hydrogen.  Below we discuss an appropriate pressure model for each category of cloud detected in [C\,{\sc ii}].

\subsection{Pressure in the CO--dark H$_2$ gas}

To use the approach in Section 4.2 to calculate the column density of CO--dark H$_2$ we need to know the thermal pressure in the different types of clouds.  For the diffuse low column density molecular clouds (no or very little CO) in the solar neighborhood, the H$_2$ layer containing C$^+$ is observed to have thermal pressures of order a few$\times$10$^3$ K cm$^{-3}$ \citep[c.f.][]{Sheffer2008,Goldsmith2013}.  Densities and temperatures, and therefore thermal pressures, have been determined from observations of absorption lines in the optical and UV for nearby  low column density clouds.  For example, \cite{Sheffer2008} analyzed UV absorption lines of H$_2$ in diffuse molecular clouds with $N$(H$_2$) column densities up to $\sim 10^{21}$\,cm$^{-2}$ (or $A_{\rm V}$ $\sim$1 mag.) and derived $T_{\rm kin}$ ranging from 50 to 120\,K, with an average of 76\,K in the molecular portion of the cloud.  \cite{Goldsmith2013} derived densities for these clouds from the UV absorption measurements of trace amounts of CO by applying an excitation model wherever reliable measurements existed for two or three of the lowest CO transitions.  \cite{Goldsmith2013} finds an  average H$_2$ density, $<$$n$(H$_2$)$>$ of 74 cm$^{-3}$ for a kinetic temperature 100\,K and 120 cm$^{-3}$ for 50\,K, corresponding to an average thermal pressure in the range 6000 to 7400 K cm$^{-3}$.  \cite{Jenkins2001} derived the thermal pressure in diffuse atomic clouds from absorption line measurements of neutral carbon towards 26 stars and find an average pressure of 2240 K cm$^{-3}$. In a later paper \cite{Jenkins2011}  extended their analysis to  89 stars and derived a mean pressure of 3800 K cm$^{-3}$ locally.  
For the dense molecular clouds, \cite{Sanders1993} estimated that the thermal pressures range from 10$^4$ to 10$^5$ K cm$^{-3}$ \citep[see also][]{Wolfire2010}. The pressures in the transition clouds likely lie in between the diffuse and dense clouds.
 
For the diffuse molecular clouds we adopt the approach taken in Paper I, which uses the Galactic pressure gradient derived by  \citet[][]{Wolfire2003} for the diffuse atomic gas, but scaled up by a factor of 1.6 to normalize to a local value intermediate between those of  \cite{Sheffer2008}, \cite{Goldsmith2013}, and \cite{Sanders1993}.  This yields $P(R_{\rm gal})=2.2\times10^4\exp(-R_{\rm gal}/5.5)$\,(K\, cm$^{-3}$), which has a value of 4700 (K cm$^{-3}$) at 8.5 kpc.  We adopt a characteristic kinetic temperature in the CO--dark H$_2$ layer of 70\,K for these clouds.  In this paper we will present our results in terms of column density, because the conversion to visual extinction depends on the metallicity which varies across the Galaxy.

The transition molecular clouds  (those with $^{12}$CO, but no $^{13}$CO) have larger column densities, hydrogen densities, and pressures than diffuse molecular clouds. The range of hydrogen column density for these clouds is fairly limited because if they get too thick or dense we should detect $^{13}$CO (see for example the models of $^{12}$CO and $^{13}$CO by  \citealt{Visser2009}).  (As we show below for the clouds analyzed here, over 70$\%$ of these clouds lie in the range  $N$(H$_2$) = 0.9 to 2.8 $\times$10$^{21}$ cm$^{-2}$, with a median value of 1.7$\times$10$^{21}$ cm$^{-2}$ ).  For lack of direct measurements of the pressure in these clouds, here we adopt a fixed pressure of 10$^4$ K cm$^{-3}$, at the lower end of the range appropriate for dense $^{13}$CO molecular clouds and at the high end of the pressure for the diffuse molecular clouds.  In Paper I we assumed a Galactic gradient for the pressure of these clouds, but over the range where most of the sources are detected it represents only a $\pm$20$\%$ difference.  We adopt a kinetic temperature of 70\,K which is consistent with the models of the CO--dark H$_2$ layer in dense molecular clouds by \cite{Wolfire2010}.  
 
\begin{figure}
 \includegraphics[width=9cm]{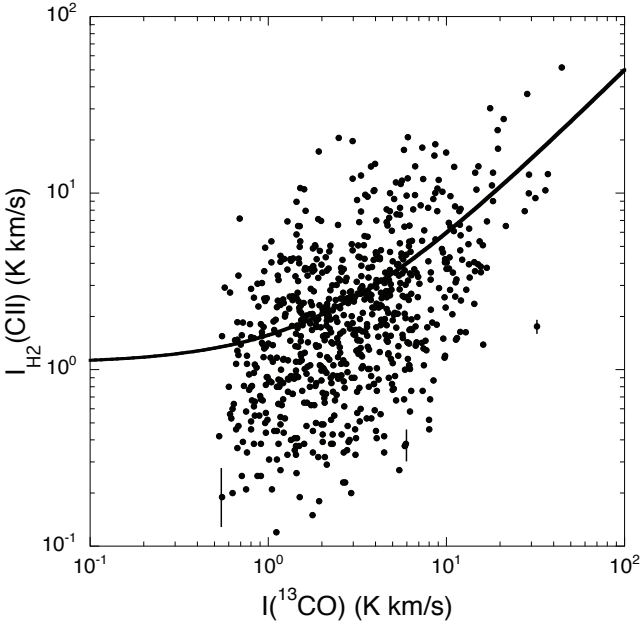}
     \caption{$I_{\rm H_{2}}$([C\,{\sc ii}]), the intensity of [C\,{\sc ii}]  in the H$_2$ layer plotted against $I$($^{13}$CO) on a log-log scale. There is a trend towards larger $I_{\rm H_{2}}$([C\,{\sc ii}])  with increasing $I$($^{13}$CO).  The solid line shows a linear fit to the data, $I_{\rm{H_2}}$([C\,{\sc ii}])= 1.08+0.49 $I$($^{13}$CO), which appears curved when transformed to the log-log plot shown here. The 1-$\sigma$ noise for $I_{\rm H_{2}}$([C\,{\sc ii}])  is indicated on three points; those for $I$($^{13}$CO) are too small to be visible on this figure and have been omitted.  It can be seen that on the scale of this figure that the S$/$N is high for most of points in this scatter plot.}
              \label{fig:fig_9_CII_H2vs13CO}
    \end{figure}

For the dense molecular clouds with $^{13}$CO the assignment of a pressure is far more difficult because the observed range of pressures in Galactic surveys is very large, 10$^4$ to 10$^5$ K cm$^{-3}$.  These clouds have a narrow shell of C$^+$, the column density of CO can be much larger, up to the upper mass limit of the Giant Molecular Clouds (GMC). Indeed the column density distribution for our sample of $^{13}$CO clouds tends to be much broader and flatter than that for the diffuse molecular and transition molecular clouds.  Examination of the relationship between the intensities of [C\,{\sc ii}] and $^{13}$CO provides a clue as to how to evaluate the variation in pressure in these clouds.  In Figure~\ref{fig:fig_9_CII_H2vs13CO} we plot the intensity of [C\,{\sc ii}]  arising from the H$_2$ gas,  $I_{\rm H_{2}}$([C\,{\sc ii}]), derived using Appendix Equation A.7 versus $I$($^{13}$CO).  Even though  there is large scatter in the data there is an overall trend for $I_{\rm H_{2}}$([C\,{\sc ii}]) to increase  with increasing $I$($^{13}$CO), especially for larger values of $I$($^{13}$CO), which correspond to the thicker and more massive clouds. A least squares linear fit to the data yields $I_{\rm{H_2}}$([C\,{\sc ii}])= 1.08+0.49 $I$($^{13}$CO) with a correlation coefficient of 0.56, which is statistically a moderately correlated relationship. 

Chemical models of the abundance profiles of C$^+$ and CO  clouds thick enough to produce detectable $^{13}$CO yield a roughly constant column density $N$(C$^+$)  whereas $N$($^{13}$CO) can have a very large range of values because C$^+$ comes from the outside layer, whereas the shielded interior where CO is found can be of any thickness within the limits of typical GMC masses and column densities per beam.  For example in the models of CO--dark H$_2$ in dense clouds by \cite{Wolfire2010} the thickness of the CO--dark H$_2$ layer varies from $N$(CO--dark H$_2$)$\sim$0.6 to 2.8 $\times$10$^{21}$ cm$^{-2}$, even as the total mass of the cloud increases from 10$^5$ to 3$\times$10$^6$ M$_{\odot}$ and the strength of the FUV radiation field changes by a factor of 10.  Thus we suggest that the large increase in  [C\,{\sc ii}] emission with increasing $^{13}$CO emission is most likely due to an increase in the hydrogen density of the [C\,{\sc ii}] emitting layer, and therefore its thermal pressure.  Assuming this relationship is correct we can use $I$($^{13}$CO) to estimate the thermal pressure in the [C\,{\sc ii}] emitting layer as follows. We use the linear relationship derived for the data in Figure~\ref{fig:fig_9_CII_H2vs13CO} to scale the thermal pressure used for modeling the dense molecular clouds traced by $^{13}$CO.  In the limit of vanishing $I$($^{13}$CO) we assume the pressure approaches that of the $^{12}$CO transition molecular clouds, assumed above to be $\sim$10$^4$ K cm$^{-3}$, therefore $P$([C\,{\sc ii}])=$10^4$[1.08+0.49$I$($^{13}$CO)] in units of K cm$^{-3}$.  For the [C\,{\sc ii}] emitting layer in $^{13}$CO clouds we adopt a kinetic temperature of $T_{\rm kin}$=70\,K, consistent with the values calculated by \cite{Wolfire2010} which vary from $\sim$50\,K to 80\,K in the CO--dark H$_2$ transition zone  (see their Figure 4). 

 
\subsection{Metallicity}

To convert $N_{\rm{H_2}}({\rm C^+})$ to $N$(H$_2$) we need to know the carbon fractional abundance  $x_{\rm H_{2}}$(C$^+$) =n(C$^+$)/n(H$_2$) across the Galaxy.   The metallicity fraction per molecular hydrogen is double that in the atomic gas, $x_{\rm H_{2}}({\rm C}^+) = 2x_{\rm H}({\rm C}^+)$.  Observations in the local interstellar medium lead to $x_{\rm H}({\rm C}^+)=n({\rm C}^+)/n({\rm H})$ in the range 1.4 to 1.8 $\times 10^{-4}$ \citep[][]{Sofia1997}, and here we adopt a value of $x_{\rm H}$(C$^+$)=1.4$\times$10$^{-4}$.  However, we need to account  for the metallicity gradients in the Galaxy.  For the carbon metallicity gradient we adopt the formula discussed 
by \cite{Wolfire2003} based on the oxygen metallicity gradient derived by \cite{Rolleston2000} over the range 3 kpc $<R_{\rm gal}<$18 kpc, and used in Paper I,
 
\begin{equation}
x_{\rm H}(C^+) = 5.51\times 10^{-4} \exp(-R_{\rm gal}/6.2). 
\end{equation}

\noindent For $R_{\rm gal}<$ 3 kpc we adopt a constant value, $x_{\rm H}({\rm C}^+) = 3.4\times 10^{-4}$ equal to that at $R_{\rm gal}$= 3 kpc.  Only a small percentage of our  [C\,{\sc ii}]  sources are at $R_{\rm gal}<$3 kpc and very few $\le$2 kpc, so any error in setting $x_{\rm H}(C^+) $ to a constant in this regime does not affect our results for the mass distributions very much.  This value is doubled in the molecular gas and $x_{\rm H_{2}}({\rm C}^+) = n({\rm C}^+)/n({\rm H}_2)$ which yields a local value near the solar system of 2.8$\times10^{-4}$. 

 
\subsection{Column Densities of the Cloud Components}

\begin{figure}
     \centering
     \includegraphics[width=8cm]{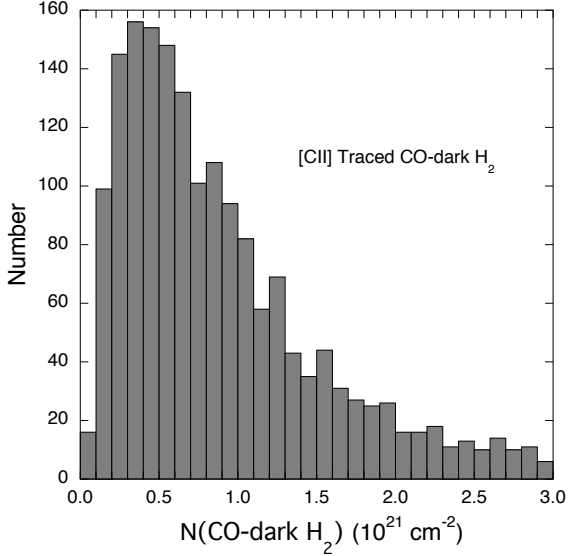}
     \caption{The distribution of the column density of the CO--dark H$_2$ gas layer in all [C\,{\sc ii}] sources in the GOT C+ inner Galaxy survey.}
         \label{fig:fig_10_NH2(dg)_all_CII}
 \end{figure}

To determine the distribution of warm CO--dark H$_2$ gas we calculate the column densities of all the cloud layers (H\,{\sc i}, C$^+$, CO) following the procedures outlined in Section 4.1 above and in Appendix A for each [C\,{\sc ii}] spectral line component. In Figure~\ref{fig:fig_10_NH2(dg)_all_CII} we plot a histogram of the column density of the CO--dark H$_2$ layer, $N$(CO--dark H$_2$) (= $N_{\rm [CII]} ({\rm H_2})$ in Equation (5)) .  For this distribution the median $N$(CO--dark H$_2$)=0.7$\times$10$^{21}$ cm$^{-2}$, the mean value is $1\times10^{21}$\,cm$^{-2}$, and there is a tail extending beyond $\sim$1.4$\times$10$^{21}$ cm$^{-2}$. About 4$\%$ of the sources lie above 2.8$\times$10$^{21}$ cm$^{-2}$. 

About 65$\%$ of the sources have a $N$(CO--dark H$_2$) less than 1$\times$10$^{21}$ cm$^{-2}$ and $\sim$90$\% \le$1.9$\times$10$^{21}$ cm$^{-2}$. These values are consistent with cloud-chemical models of  low column density clouds \citep[][]{Visser2009} and the edges of more massive clouds \citep[][]{Wolfire2010}. However, roughly 10$\%$ of the clouds in Figure~\ref{fig:fig_10_NH2(dg)_all_CII} have larger $N$(CO--dark H$_2$). Such thick layers of CO--dark H$_2$ are not expected theoretically.  It is possible that these large values result from underestimating the true pressure in the H$_2$ gas or arise from several lower column density clouds blended into one broad spectral feature.

\begin{figure}
     \centering
     \includegraphics[width=8cm]{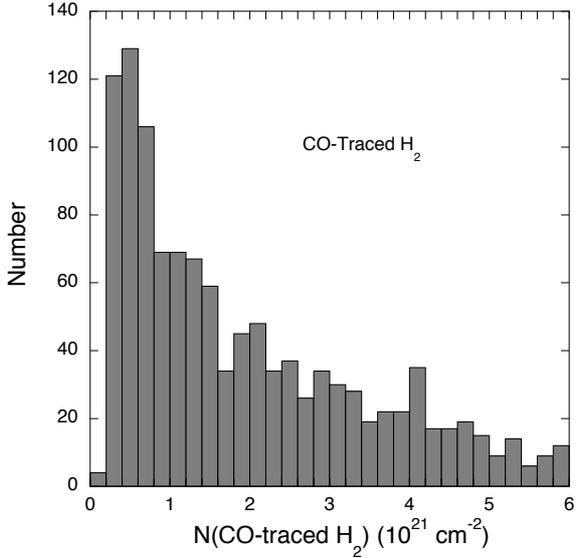}
     \caption{The distribution of the column density of the CO-traced gas, N(CO--traced H$_2$), in the [C\,{\sc ii}] sources.}
         \label{fig:fig_11_NH2(CO)_CII_clouds}
 \end{figure}

In Figure~\ref{fig:fig_11_NH2(CO)_CII_clouds}, we plot the distribution of column densities of the CO--traced H$_2$, $N$(CO--traced H$_2$).  There are $\sim$1250 CO sources in this figure, less than the number of C$^+$ sources because a fraction of our [C\,{\sc ii}]  comes from clouds without CO.  The median and mean $N$(CO--traced H$_2$) are 1.5 and 2.3 $\times$10$^{21}$ cm$^{-2}$, respectively. Over 80$\%$  of the clouds have a low CO--traced H$_2$ column density, with $N$(CO--traced H$_2$)$\le$3.8$\times$10$^{21}$ cm$^{-2}$, and $\sim$90$\%$ have $\le$5.6$\times$10$^{21}$ cm$^{-2}$.  However, unlike the case for the CO--dark H$_2$ layer, there is no restriction on how thick the dense CO--traced H$_2$ interior can be, within the limits of typical GMC masses, and about 3$\%$ of the clouds have $N$(CO--traced H$_2$)$>$9$\times$10$^{21}$ cm$^{-2}$ (not shown in Figure~\ref{fig:fig_11_NH2(CO)_CII_clouds}).

\begin{figure}
     \centering
     \includegraphics[width=6.95cm]{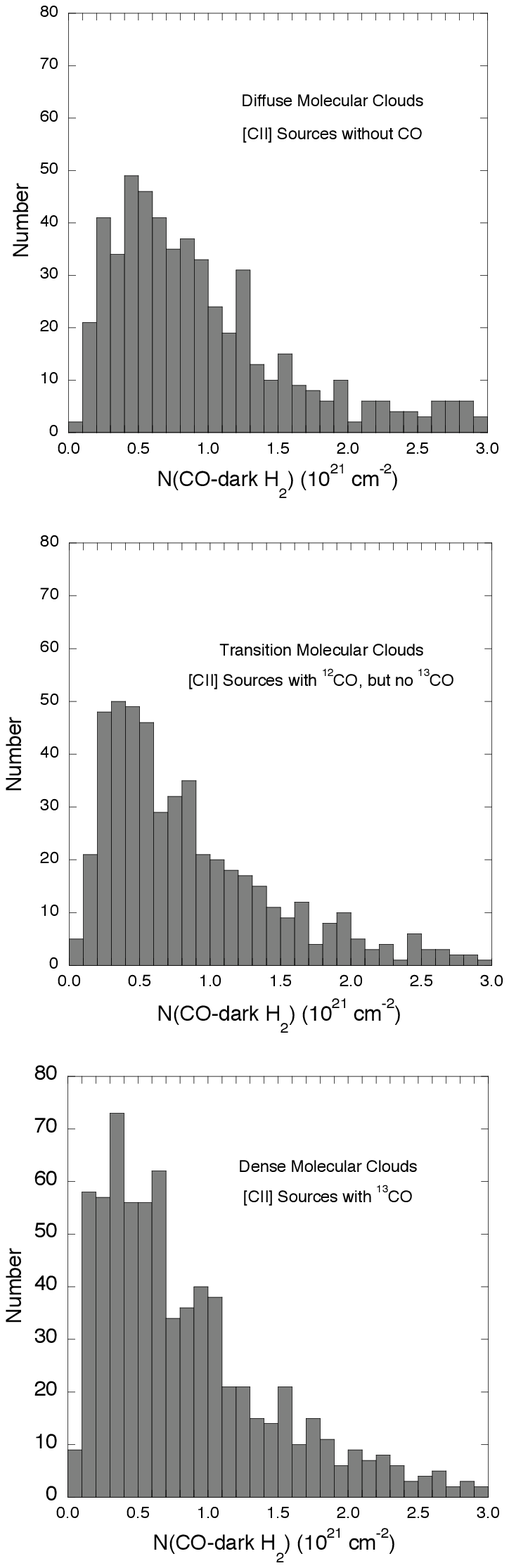}
     \caption{The distribution of the column density of the CO--dark H$_2$ gas layer sorted by type of cloud in the GOT C+ inner Galaxy survey: (top) diffuse molecular clouds; (middle) transition molecular clouds; and, (bottom) dense molecular clouds.}
         \label{fig:fig_12_NH2(dg)_vs_CxO}
 \end{figure}

In Figure~\ref{fig:fig_12_NH2(dg)_vs_CxO} we compare the column density of the CO--dark H$_2$ layers in three types of clouds: diffuse molecular clouds without CO, transition molecular clouds with $^{12}$CO but no $^{13}$CO, and dense molecular clouds with $^{13}$CO (some of which also contain C$^{18}$O). The distributions of $N$(CO--dark H$_2$) are similar for all three types of clouds and they have comparable median (0.79, 0.69, and 0.65)$\times$10$^{21}$ cm$^{-2}$ and mean values (1.0, 0.94, and 0.93)$\times$10$^{21}$ cm$^{-2}$.  (For reference note that the mean column density for our sample of sources without CO corresponds to $A_{\rm V}\sim$ 1 mag., using the standard conversion to visual extinction in the solar neighborhood.)  The mean value of $N$(CO--dark H$_2$) in the dense molecular clouds is consistent with models of this layer by \cite{Wolfire2010}, however, they found a nearly constant thickness of $N$(CO--dark H$_2$) over a wide range of conditions for the dense clouds,  whereas we see significant variability. In the next Section we make a more detailed comparison of observations with models. 
 
\begin{figure*}[t]
     \centering
     \includegraphics[scale=1.0]{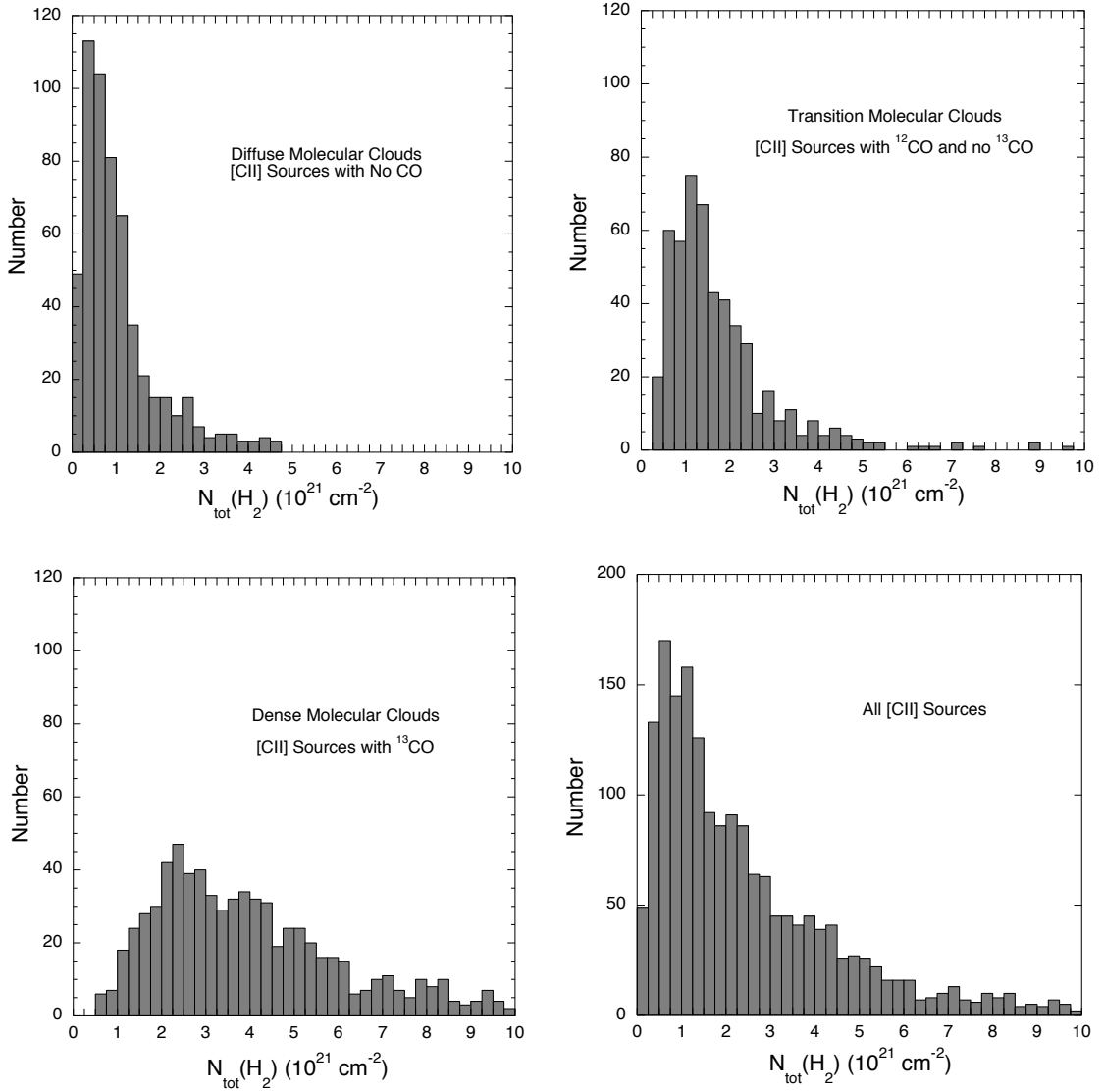}
     \caption{The distribution of total H$_2$ column density for different types of clouds: (top left) diffuse molecular clouds with no CO, (top right) $^{12}$CO  transition molecular clouds, (bottom left) dense molecular clouds with $^{13}$CO, and (bottom right) all [C\,{\sc ii}]  sources.  Note that the vertical scale is different in the last panel.} 
     \label{fig:fig_13_Ntot(H2)_vs_CxO}
\end{figure*}

Figure~\ref{fig:fig_13_Ntot(H2)_vs_CxO}  shows the total H$_2$ column density, $N_{\rm tot}$(H$_2$)=$N$(CO--dark H$_2$)+$N$(CO--traced H$_2$),  for different categories of clouds.  The addition of the H\,{\sc i} contribution to  the column density does not change the distribution much as $N$(H\,{\sc i})  is generally small compared to the other terms in the clouds detected in [C\,{\sc ii}].  The peak in the distribution of the total H$_2$ column density in diffuse molecular and transition molecular clouds is narrow, whereas that in  the dense molecular clouds is broader.  The median total H$_2$ column density is 0.8$\times$10$^{21}$ cm$^{-2}$ for the diffuse molecular clouds, 1.4$\times$10$^{21}$ cm$^{-2}$ for the transition clouds, and 3.7$\times$10$^{21}$ cm$^{-2}$ dense clouds.  The breadth of the distribution in the total H$_2$ column density is indicated by the mean values and standard deviation of $N_{\rm tot}$(H$_2$), for these three categories:  (1.1$\pm$0.9)$\times$10$^{21}$ cm$^{-2}$ for diffuse molecular clouds, (1.8$\pm$1.3)$\times$10$^{21}$ cm$^{-2}$ for the transition molecular clouds, and (4.4$\pm$2.9)$\times$10$^{21}$ cm$^{-2}$ for the dense molecular clouds. The distribution for all the [C\,{\sc ii}] sources combined is given in the lower-right panel of Figure~\ref{fig:fig_13_Ntot(H2)_vs_CxO}. This distribution has a median $N_{\rm tot}$(H$_2$) = 1.8$\times$10$^{21}$ cm$^{-2}$, and a mean value of (2.6$\pm$2.5)$\times$10$^{21}$ cm$^{-2}$.  Overall, about 80$\%$ of the sources have a total molecular hydrogen column density less than  4$\times$10$^{21}$ cm$^{-2}$ and $\sim$90$\%$ have total H$_2$ column densities $\le$5$\times$10$^{21}$ cm$^{-2}$.  Fewer than 2$\%$ of the sources have large column densities, $\ge$10$\times$10$^{21}$ cm$^{-2}$, most of which are detected in C$^{18}$O; these could be dense PDRs.   About 30$\%$ of the sources have a low total H$_2$ column density $\le$1$\times$10$^{21}$ cm$^{-2}$.   

\subsection{CO--dark H$_2$ Mass Distribution}

\begin{figure*}[t]
     \centering
       \includegraphics[scale=1.0]{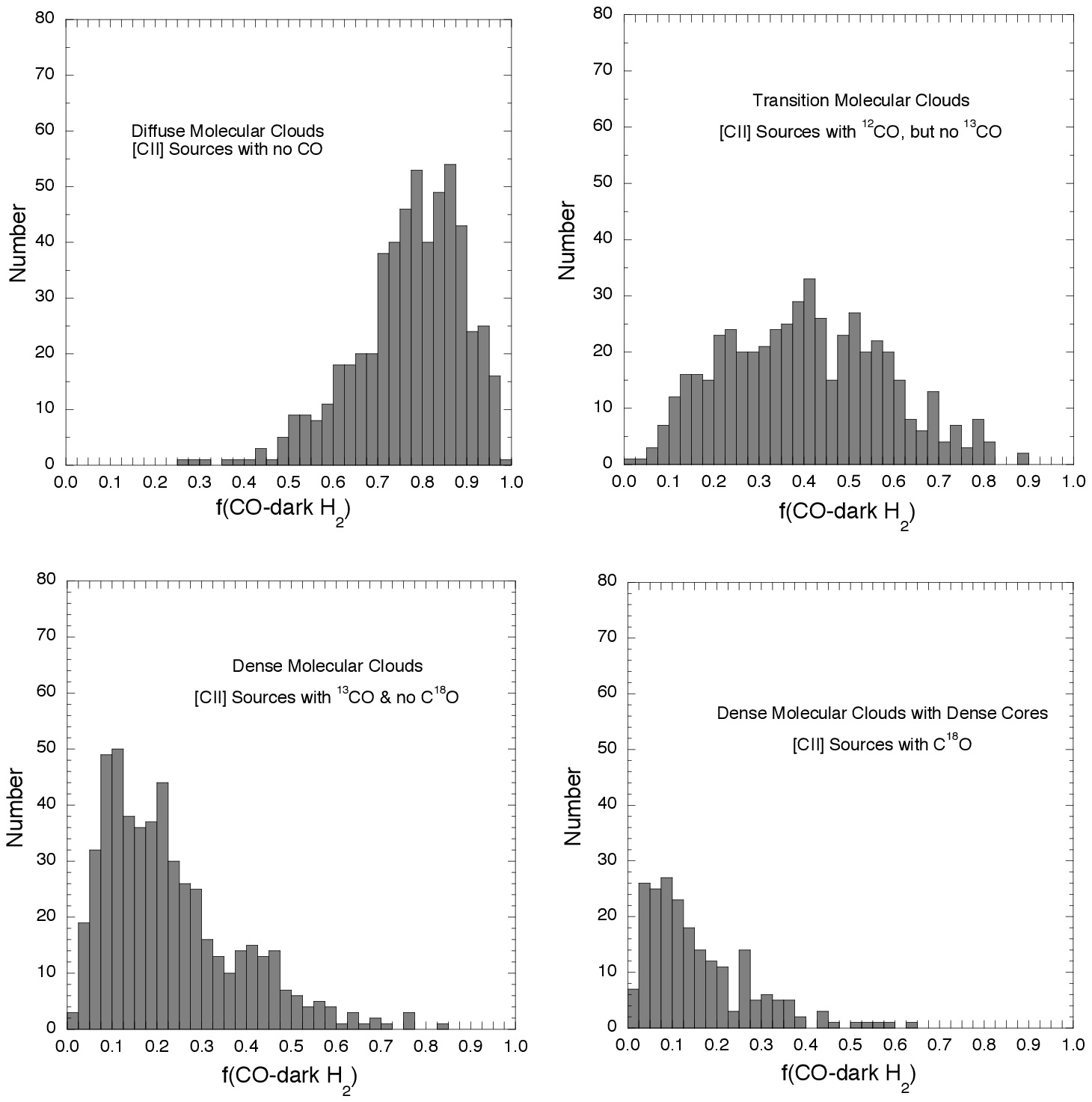}
     \caption{The fraction by mass of CO--dark H$_2$ gas for different classes of clouds: (upper left)  diffuse molecular clouds, the [C\,{\sc ii}]  sources with no CO; (upper right) transition molecular clouds, [C\,{\sc ii}]  sources with only $^{12}$CO; (lower left) dense molecular clouds, with $^{13}$CO and no C$^{18}$O; and, (lower right) dense molecular clouds with $^{13}$CO and a dense C$^{18}$O core (or cores).}
          \label{fig:fig_14_f(dg)_CII_vs_yCxO}
 \end{figure*}

 We use the column densities to calculate the fractional mass, $f$, in the CO--dark H$_2$ gas in each cloud using, 

\begin{equation}
\quad
f= \frac{2N(\rm CO{-}dark\, H_2)}{N({\rm HI})+2N({\rm CO{-}dark\, H_2})+2N({\rm CO{-}traced\, H_2})}\,
\end{equation}

\noindent where the factors of two account for the mass of H$_2$ compared to atomic hydrogen. Unlike the analysis in Paper I, which derived the fraction of CO--dark H$_2$ gas in the Galactic ISM, here we calculate only the fractional mass of CO--dark H$_2$ in each cloud  and how the distribution of $f$ might differ among the various categories of clouds. In Figure~\ref{fig:fig_14_f(dg)_CII_vs_yCxO}  we plot the distribution of $f$ for: diffuse molecular hydrogen (no CO), transition molecular clouds, and dense molecular clouds with $^{13}$CO but no C$^{18}$O, and dense molecular clouds with C$^{18}$O).  As expected the diffuse molecular clouds have the largest fraction of CO--dark H$_2$ by mass because they do not have extensive molecular cores.  In contrast, the dense molecular clouds, as traced by $^{13}$CO, have significant molecular gas in their shielded cores and their CO--dark H$_2$ mass fraction is much smaller.  The dense C$^{18}$O clouds tend to have the largest CO column densities and so the smallest fraction of CO--dark H$_2$. The $^{12}$CO transition molecular clouds lie somewhere in between and have a very broad range of $f$. 

\begin{figure}
     \centering
     \includegraphics[width=8cm]{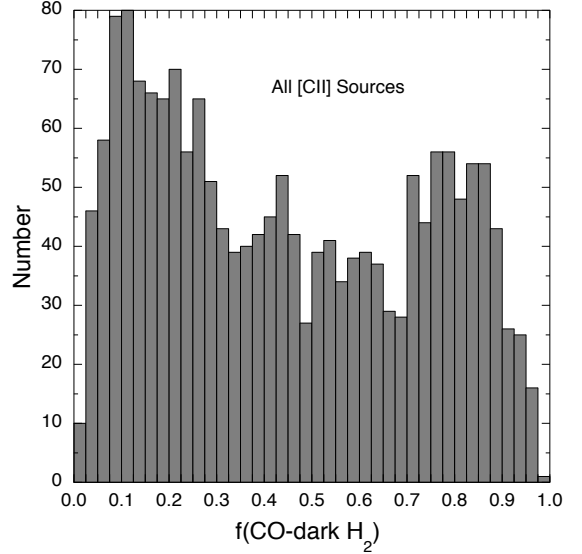}
     \caption{The fraction by mass of CO--dark H$_2$ gas for all the [C\,{\sc ii}]  sources.}
         \label{fig:fig_15_f(dg)_all_CII}
 \end{figure}

In Figure~\ref{fig:fig_15_f(dg)_all_CII}   we plot the distribution of fractional mass in the CO--dark H$_2$ gas layer for all the [C\,{\sc ii}] clouds. The two peaks correspond to  diffuse molecular and dense molecular clouds.  In Table~\ref{tab:f_CO_darkgas} we list the median of the fractional mass of CO--dark H$_2$ gas, its mean value $<$$f$$>$, and 1--$\sigma$ standard deviation. 

\begin{table}
\center
\caption{Mass fraction of CO--dark gas}
\label{tab:f_CO_darkgas}
\begin{tabular}{c c c c}
\hline\hline
 Cloud Category & $f-$median & $<$$f$$> \pm 1\sigma$ & Tracer \\
   \hline
Diffuse Molecular & 0.78  & 0.77 $\pm$ 0.12 & No CO  \\
Transition Molecular  & 0.42 & 0.43 $\pm$ 0.18 & $^{12}$CO, no $^{13}$CO \\
Dense Molecular & 0.18 &  0.21 $\pm$ 0.15  & $^{13}$CO \\
Dense Cores &  0.12 & 0.16 $\pm$ 0.12  & C$^{18}$O\\
\hline
All [C\,{\sc ii}] Sources  & 0.41 &  0.44 $\pm$ 0.28  &  \\
\hline
\end{tabular}\\
\end{table}

The diffuse molecular clouds detected in the GOT C+ survey have significant fractions of CO--dark H$_2$ gas, typically more than 70 percent.  The fact that so few purely atomic H\,{\sc i} diffuse clouds are present in the data set (i.e. those with little or no H$_2$) is due to a lack of sensitivity to [C\,{\sc ii}] for low H\,{\sc i} column and/or low volume density clouds.  For the transition molecular clouds the distribution in $f$(CO--dark H$_2$) is shifted towards lower values and $<$$f$$>$$\sim$0.4.  This result is consistent with our picture of the profile of carbon in clouds: it is primarily ionized in the envelope and makes a somewhat sharp transition to CO (with a narrow layer mixing C$^+$, C, and CO) as the FUV radiation is attenuated by dust, H$_2$ lines, and CO line self-shielding.   In the dense molecular clouds where $^{13}$CO is detected, the CO molecular gas region is even thicker and the fraction of CO--dark H$_2$ becomes smaller, with $<$$f$$>$$\sim$0.2.  Taken together the ensemble of [C\,{\sc ii}] clouds have a mass fraction of CO--dark H$_2$, $<$$f$$>$$\sim$0.44.  These results omit the contribution of CO clouds where we do not detect [C\,{\sc ii}].  In Paper I the contributions of all the [C\,{\sc ii}] and CO gas is taken into account in deriving a Galactic radial solution to the fraction of CO--dark H$_2$, and yields a smaller fraction, $<f>\sim 30\%$ averaged over the entire Galaxy.

 
\section{Comparison to cloud chemical models}

Over the past forty years ever more sophisticated models have been constructed of PDRs taking into account the many physical and chemical processes that determine their structure.  In interstellar clouds photodissociation of CO by UV is the most important destruction mechanism and photoionization keeps carbon ionized.  Hence the treatment of the photodestruction processes and the corresponding radiative transport in the cloud is key to modeling the C$^+$ to CO transition and calculating the thickness of the CO--dark H$_2$ layer. Here we compare our results to the recent cloud chemical models of the transition from C$^+$ to CO by  \cite{Visser2009} and \cite{Wolfire2010}.  The depth of the transition from C$^+$ to CO yields the column density of the CO--dark H$_2$ layer. 

The models of \cite{Visser2009} were aimed at understanding the depth dependence of the onset of the formation of CO and its isotopologues under a variety of cloud conditions. They used a semi-infinite slab model and integrate step-wise into the clouds to calculate the UV radiation field.  Attenuation of the radiation includes continuum absorption by dust, shielding by H$_2$ and CO, and self-shielding of CO isotopologues using the latest laboratory and theoretical values for the line parameters.  They adopted the benchmark chemical network and chemical reaction rates from a PDR benchmark study \citep{Rollig2007}.  This reaction network incorporates  31 atomic and molecular species consisting only of H, He, C, O, and electrons, however they add the $^{13}$C, $^{17}$O, and $^{18}$O isotopes and the corresponding isotope exchange reactions.  The addition of isotopes of carbon and oxygen increases the number of species to 118 and the chemical reaction network to 1723. The restriction to 118 species focus on the key carbon network producing CO, while keeping the number of reactions manageable for a parameter study. (Note, some very large networks and data bases contain over 400 species and over 4500 reactions \cite[c.f.][]{McElroy2013} without isotopes, so any full network chemistry including isotopologues would be much more computationally demanding for modeling the cloud chemistry over a large parameter space.)  The restricted network does leave out potentially important species for the ionization balance, such as sulfur \cite[c.f.][Section 3.1.2]{Rollig2007}.

\cite{Visser2009} calculate the abundances and column densities of the CO isotopologues in diffuse and transition molecular clouds (A$_{\rm v}$$\le$4 mag.\,) for a range of physical parameters.  They calculated $N({\rm CO})$ as a function of $N$(H$_2$) for a grid covering the following range of parameters: $n$(H$_2$)=100 -- 1000 cm$^{-3}$, $T_{\rm gas}$ = 15 -- 100\,K, and FUV strength $\chi$ = (1 -- 10)$\chi_{\rm 0}$, where $\chi_{\rm 0}$ is the local value in the solar neighborhood,  for about 60 model runs.  They plotted the results for all the depth steps in their grid for all the parameters $n$(H$_2$), $T_{\rm gas}$, and $\chi$.   The output consists of thousands of model values for the column density of the CO isotopologues as a function of $N$(H$_2$). Details of their model results are presented in figures and tables both in their text and on-line.  Figure 7 in \cite{Visser2009} displays $N$($^{12}$CO) versus $N$(H$_2$) for all points with $A_V<$ 4 mag.\, from their grid of cloud models.  The scatter in the model results is due to the different physical parameters assumed for the cloud models. In addition to the set of maximum and minimum column densities of $N$(CO), they also generate a 100-point mean of the model results by averaging the solutions in fixed grid regions at over 30 depths into the cloud to facilitate comparison to observations. They found that the transition from a C$^+$ to a CO dominated regime can occur over a wide range of $N$(H$_2$) from $\sim$a few$\times$10$^{20}$ to $\sim$2$\times$10$^{21}$ cm$^{-2}$ (see their Figure 7).  

\cite{Wolfire2010}  modeled the thickness of the CO--dark H$_2$ layer in a spherical model of massive dense molecular clouds and considered the effects of the strength of the FUV field, metallicity, density, and cloud mass.  They found that for $\chi$ in the range (3 --  30)$\chi_{\rm o}$ the CO--dark H$_2$ column density, which they measured in $A_{\rm V}$, varies very little.  However, they did find that the solutions depended on the metallicity (see their Figure 10).

\begin{figure}
     \centering
     \includegraphics[width=8cm]{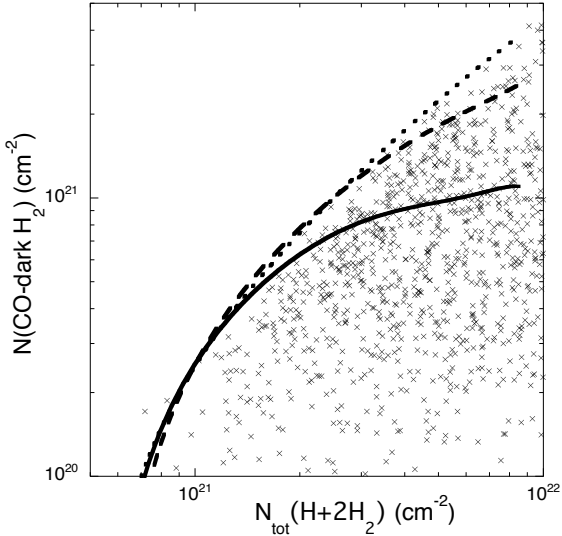}
     \caption{The column density of CO--dark H$_2$ gas for all CO sources as a function of the total hydrogen column density $N_{\rm tot}$=$N$(H\,{\sc i})+2$N$(H$_2$).  The curves are $N$(CO--dark H$_2$) calculated from the models of \cite{Visser2009} as discussed in the text.  The central curve (dashed line) is the 100-point means of their model shown in the right panel of their Figure 7. The top curve (dotted line) is the lower limit on the solutions for $N$(CO), which yields a maximum for $N$(CO--dark H$_2$), and the bottom curve (solid line) is the upper limit on $N$(CO), which yields a minimum for $N$(CO--dark H$_2$), as a function of $N$(H$_2$) for the range of parameter space used in their models.}
         \label{fig:fig_16_N(dg)_Ntot_Model}
 \end{figure}

In Figure~\ref{fig:fig_16_N(dg)_Ntot_Model} we plot $N$(CO--dark H$_2$) as a function of the total hydrogen column density, $N_{\rm tot}$=$N$(H\,{\sc i})+2$N$(H$_2$), for all clouds detected in CO in our survey up to $N_{\rm tot}$=10$^{22}$ cm$^{-2}$. For comparison we have plotted some of the results from \cite{Visser2009} in Figure~\ref{fig:fig_16_N(dg)_Ntot_Model}.  We have used the solutions for $N$(CO) versus $N$(H$_2$) in their Figure 7 and the Tables  to calculate the upper and lower bounds on $N$(CO--dark H$_2$) and for their 100--point mean solution.  To calculate $N$(CO--dark H$_2$) from \cite{Visser2009} we first calculate the total carbon column density, $N_{\rm tot}$(C)=x$_H$(C$)N_{\rm tot}$, and then $N$(C$^+$)=$N_{\rm tot}$(C)-$N$(CO), and finally $N$(CO--dark H$_2$)=$N$(C$^+$)$/x_{\rm H_2}({\rm C^+}$), where we neglect the small contribution from $N$(H\,{\sc i}).

From these column densities we calculate the boundary of the maximum, minimum, and 100-point mean fraction $N$(CO--dark H$_2$) and plot them in  Figure~\ref{fig:fig_16_N(dg)_Ntot_Model}.  While many data points fall within the bounds of their models, the majority of the clouds have much less $N$(CO--dark H$_2$) than the models.  However, in comparing our values of $N$(CO--dark H$_2$) to the model values it must be kept in mind that \cite{Visser2009} assume a metallicity appropriate to the solar neighborhood which does not apply  to most of the sources in our survey.  In the work of \cite{Wolfire2010} on much more massive clouds they found that an increase in metallicity resulted in a lower column density of CO--dark H$_2$ all other factors kept equal. There is no comparable set of models with different metallicity along the lines of \cite{Visser2009} available for more detailed comparison.

In Figure~\ref{fig:fig_17_f(dg)_Ntot_Model} we plot $f$(CO--dark H$_2$) as a function of the total hydrogen column density, $N_{\rm tot}$=$N$(H\,{\sc i})+2$N$(H$_2$), for all 1804 sources, separated into the three cloud categories: diffuse molecular (red $\circ$), $^{12}$CO transition clouds (blue $\times$), and the dense $^{13}$CO clouds ({black $\bigtriangleup$).  The sources cluster in regions as expected with the diffuse clouds having the highest $f$(CO--dark H$_2$) on average and the dense clouds the lowest values, corresponding to the distributions seen in Figure~\ref{fig:fig_14_f(dg)_CII_vs_yCxO}. For comparison we again use the results from \cite{Visser2009} and include those from \cite{Wolfire2010} in Figure~\ref{fig:fig_17_f(dg)_Ntot_Model}.  The figure shows just how large and variable  are the model $f$(CO--dark H$_2$).  As can be seen only about half the GOT C+ sources fall within the maximum and minimum curves from \cite{Visser2009}, but this result is not unexpected as at higher metallicity  $N$(CO-dark H$_2$) will be lower for a given N$_{\rm tot}$, as discussed for Figure~\ref{fig:fig_16_N(dg)_Ntot_Model}.  For example at R$_{\rm gal}$$\sim$4 kpc the metallicity is double the solar value, and curves for $f$(CO--dark H$_2$) will be approximately proportionally lower. Furthermore the \cite{Visser2009} models are limited to a visual extinction $\sim$4 mag.\, so do not model massive clouds with large column densities. In contrast \cite{Wolfire2010} only model very massive clouds with high average column densities corresponding to $N$(H$_2$) = 7.5 and 15$\times$10$^{21}$ cm$^{-2}$, but they do consider the dependence of $f$(CO--dark H$_2$) on metallicity, where $Z^{\prime}=Z/Z_\odot$ is the metallicity with respect to the solar value. 
 
 In Figure~\ref{fig:fig_17_f(dg)_Ntot_Model} we also plot the solutions for $f$(CO--dark H$_2$) from \cite{Wolfire2010} for three values of the metallicity with respect to the solar value, $Z^{\prime}$ = 0.5 (red), 1.0 (blue), and 1.9 (black) times $Z_\odot$, where $Z_{\odot}$ is the solar value.  These values correspond to Galactic radii, R$_{\rm gal}$= 12.8, 8.5, and 4.5 kpc, respectively.  For each value of $Z^\prime$ there are two model points corresponding to N$_{\rm tot}$ = 1.5$\times$10$^{22}$ and 3$\times$10$^{22}$ cm$^{-2}$ (or $N$(H$_2$)= 7.5 and 15 $\times$10$^{21}$ cm$^{-2}$ in \citealt{Wolfire2010}). We connect these points with a linear fit in Figure~\ref{fig:fig_17_f(dg)_Ntot_Model}. For the enhanced metallicity, $f$(CO--dark H$_2$) is lower, ranging from 0.25 to 0.17.   Here again our values for  $f$(CO--dark H$_2$) are generally smaller than that modeled by \cite{Wolfire2010} for the massive clouds with large column densities. However, the median value of $f$(CO--dark H$_2$) for the GOT C+ data base of dense $^{13}$CO clouds is 0.18 (Table~\ref{tab:f_CO_darkgas}) in rough agreement with the values from \cite{Wolfire2010}.
 
  \begin{figure}
     \centering
     \includegraphics[width=8cm]{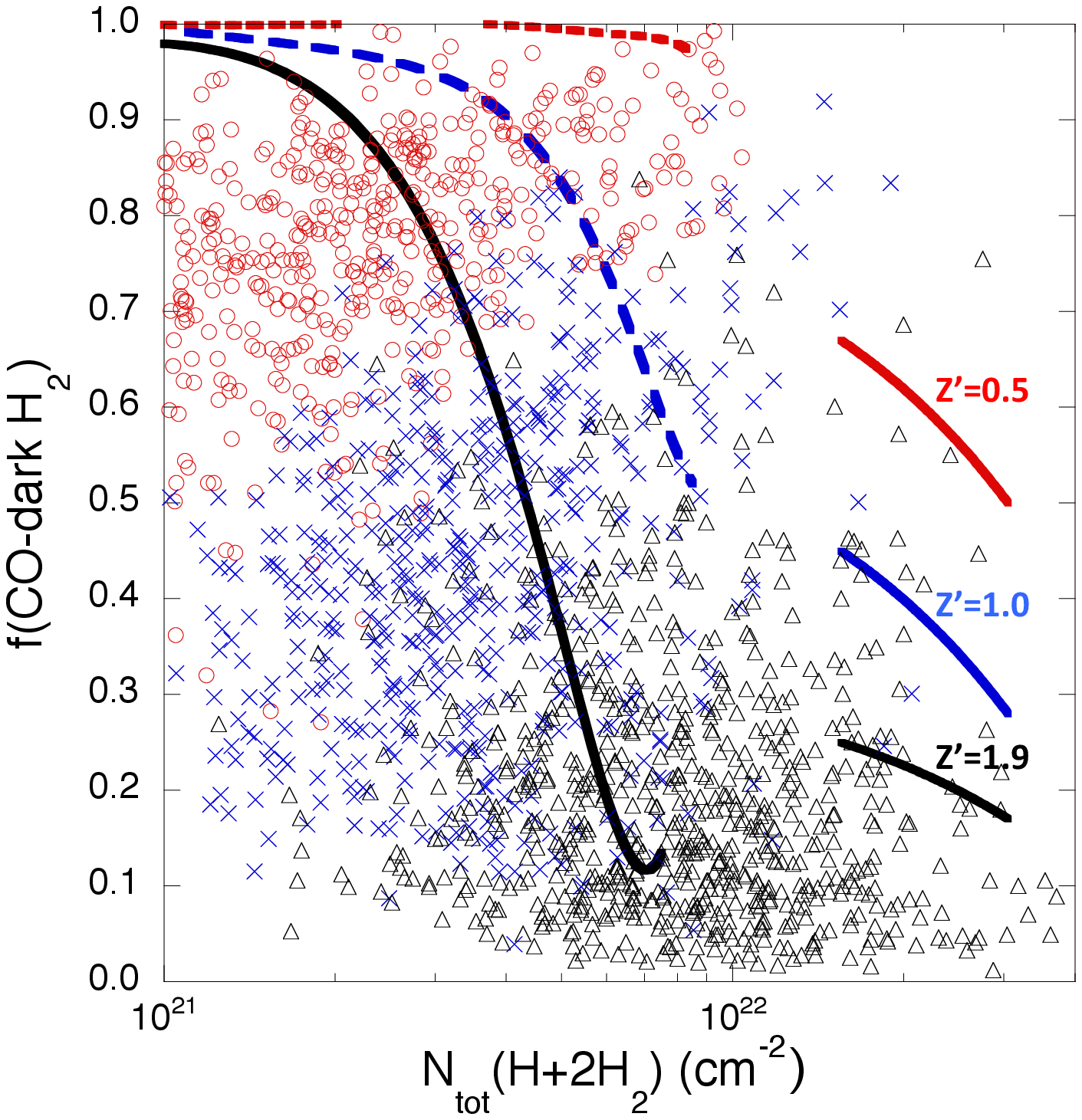}
     \caption{The fraction by mass of CO--dark H$_2$ gas for all sources as a function of the total hydrogen column density $N_{\rm tot}$=$N$(H\,{\sc i})+2$N$(H$_2$):  diffuse molecular clouds (red $\circ$), $^{12}$CO transition molecular clouds (blue $\times$), and $^{13}$CO dense molecular clouds (black $\bigtriangleup$).  The curves on the left are $f$(CO--dark H$_2$) calculated from the models of \cite{Visser2009} as discussed in the text.  The central curve (blue dashed line) is the 100-point means of their model shown in the right panel of their Figure 7. The top curve (red dotted line) is the lower limit on the solutions for $N$(CO), while the bottom curve (solid black line) is the upper limit on $N$(CO) as a function of $N$(H$_2$) for the range of parameter space used in their models.  On the right side we plot the few values of $f$(CO--dark H$_2$) calculated by \cite{Wolfire2010} for massive clouds for different values of metallicity relative to the solar value, Z$^\prime$=0.5 (red line), 1.0 (blue line), and 1.9 (black line).}
         \label{fig:fig_17_f(dg)_Ntot_Model}
 \end{figure}

As seen above, lower metallicity results in clouds with larger $f$(CO--dark H$_2$) so we expect that there should be a Galactic gradient in $f$(CO--dark H$_2$). In Paper I we found that the azimuthally averaged fraction of CO--dark H$_2$ increases with Galactocentric distance, and we get a similar trend using the cloud components.   We divided the Galaxy in 0.5 kpc rings and summed all the H\,{\sc i}, CO--traced H$_2$ and CO--dark H$_2$ column densities within each ring from sources with [C\,{\sc ii}]. Including the CO sources without [C\,{\sc ii}] introduces an uncertainty in how much hidden CO--dark H$_2$ there is in these clouds.  We used these values to calculate the average $f$(CO--dark H$_2$) within each 0.5 kpc thick ring and plot the results in Figure~\ref{fig:fig_18_f(dg)_sum_vs_R}. We see a systematic trend for $f$ from about 0.2 in the inner Galaxy to $\sim$0.55 at R$_{\odot}$.  This trend is consistent with the results expected for a metallicity gradient and because there are relatively more diffuse molecular clouds, with higher $f$(CO--dark H$_2$) values, approaching R$_{\odot}$ and relatively more massive clouds, as indicated by the C$^{18}$O clouds in the interior (see lower right panel in Figure~\ref{fig:fig_4_CII_R_CO_type}, with much smaller $f$(CO--dark H$_2$).  While the exact values of $f$(CO--dark H$_2$) in this figure may be slightly uncertain because we cannot calculate the corresponding CO--dark H$_2$ for the clouds seen in CO but not [C\,{\sc ii}], we expect  the overall trend to be valid.

\begin{figure}
     \centering
     \includegraphics[width=8cm]{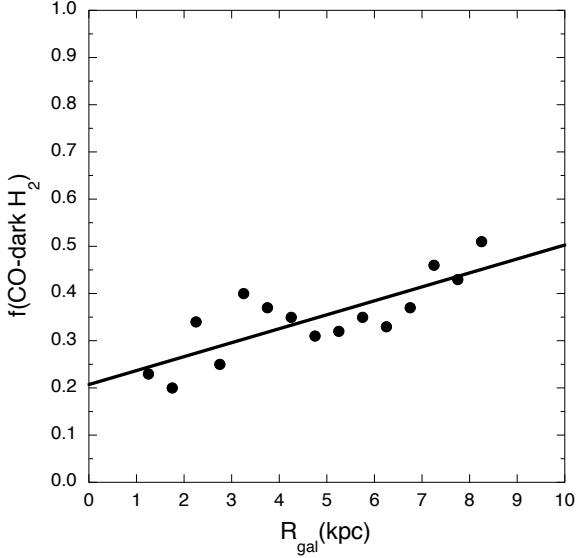}
    \caption{The distribution of $f$(CO--dark H$_2$) in CO sources with detectable [C\,{\sc ii}] averaged over rings as a function of Galactic radius.}
         \label{fig:fig_18_f(dg)_sum_vs_R}
 \end{figure}
 


\section{Summary}
\label{sec:summary}

The GOT C+ survey is the first step in providing a velocity resolved map of the [C\,{\sc ii}] in the Galactic disk.  In Paper I we discussed the global distribution of H\,{\sc i}, CO, and [C\,{\sc ii}]  and derived the azimuthally averaged distribution of ISM gas components for $b$=0$\degr$.  
In this paper we have used the spectral information to extract individual cloud features in [C\,{\sc ii}]  and characterized statistically the properties of clouds as labeled by whether they also contained $^{12}$CO and $^{13}$CO.  The approaches used in Paper I and the present work complement each other.  While Paper I accounts for all the [C\,{\sc ii}] intensity detected in the survey, the present work accounts for only about 50$\%$, the fraction of the [C\,{\sc ii}] intensity identified in narrow line ($\Delta$V$\le$8 \kmsnospace) clouds.  Thus here we provide a better picture of the global cloud properties while Paper I provides a better picture of the Galactic global [C\,{\sc ii}] distribution of each ISM component.

We identified over 1800 narrow line [C\,{\sc ii}] features in the inner Galaxy bounded by {\it l} =$270\degr$ to $57\degr$, {\it b}$\le \pm1\degr$, and $R_{\rm gal} \le 9$\,kpc.  These [C\,{\sc ii}] features were subdivided into three categories of clouds depending on the presence of CO isotopologues as follows: 557 diffuse molecular clouds with no CO; 513 transition molecular clouds with $^{12}$CO, but no $^{13}$CO; and, 734 dense molecular clouds with $^{13}$CO (212 of these dense clouds also had C$^{18}$O).  In addition, we identified 714 CO emitting clouds with no detectable [C\,{\sc ii}] at the level of the sensitivity of the GOT C+ survey. Thus it appears that our  [C\,{\sc ii}] survey is sampling a good distribution of clouds at different evolutionary stages across much of the Galactic disk. The distribution of clouds in all [C\,{\sc ii}] categories is roughly flat between 4 and 7 kpc, but the CO clouds without [C\,{\sc ii}] have a peak around 8 kpc.  

We used our   [C\,{\sc ii}] data base along with CO isotopologue and H\,{\sc i} observations to calculate the gas column density of  different layers of the cloud: diffuse H\,{\sc i} envelope, warm CO--dark H$_2$ envelope, and CO interior.  In the case of  [C\,{\sc ii}] there is only one transition to work with and so, to estimate the density and temperature in the [C\,{\sc ii}] emitting layer, we have adopted a constant temperature in the layer and use thermal pressure models to determine the likely hydrogen density.  The thermal pressure of the layer varies in a systematic way (on average) as one progresses from low density diffuse molecular clouds to the dense molecular clouds.  We find that the mean column density of CO--dark H$_2$ is similar for the three types of clouds with  $N_{\rm tot}$  = 1.0$\times$10$^{21}$ cm$^{-2}$, 0.94$\times$10$^{21}$cm$^{-2}$, and 0.93$\times$10$^{21}$ cm$^{-2}$ as one progresses from diffuse molecular, through CO transition molecular, to dense $^{13}$CO molecular clouds. 

We calculated the fractional CO--dark H$_2$ mass in each cloud and find that it decreases from an average of $\sim$77$\%$ in the diffuse molecular clouds, with no CO interiors, to as little as $\sim$20$\%$ in the dense molecular clouds with $^{13}$CO cores. The fraction in diffuse and transition molecular clouds is in general less than that calculated in the \cite{Visser2009} models, but this might be due to the choice of metallicity adopted in their models which is lower than that where most of our sources are located in the Galaxy. The fraction in the dense molecular clouds is $\sim$21$\%$ compared to the roughly 30$\%$ predicted in the dense cloud models of \cite{Wolfire2010}. While the mean column densities and $f$(CO--dark H$_2$) are roughly  in agreement with the models of \cite{Wolfire2010}, we find more dispersion in the actual data than predicted by their models.  This dispersion could be due to the application of our thermal pressure model to an ensemble of clouds, rather than knowing the exact pressure for each cloud, or it could be real, but it will take additional probes of the pressure to decide. We conclude that a significant fraction of the mass of these clouds is in the CO--dark H$_2$ gas traced by [C\,{\sc ii}] 158{\micron} emission.   

The largest uncertainty in our estimates of the amount of material seen in [C\,{\sc ii}] is the lack of precise knowledge of the temperature and density (or pressure) in the CO--dark H$_2$ layer.   We have estimated these parameters from observations and models of the thermal pressure in clouds.  A more direct approach would be to use the two fine-structure lines of [C\,{\sc i}] to constrain the density and temperature at the boundary where [C\,{\sc ii}]  is converted to [C\,{\sc i}] and CO.  Currently we do not have such a data base available for the GOT C+ survey. 

Another uncertainty in deriving our column densities from [C\,{\sc ii}] is the assumption of the homogeneity of the clouds, which can be either uniform low density envelopes or high density clumps with a low filling factor (the emission rate in the clump model will be much higher per unit volume than in the uniform model as it is $\propto n(\rm H_2)^2$ and results in a lower column density).  Finally, we have assumed that we can compare our observationally derived column densities to steady state models of clouds, but we could be observing clouds in different stages of evolution.  Time dependent model calculations show that at early times a large C$^+$ envelope can be maintained, much larger than the final steady state layer \citep[c.f.][]{Lee1996}.  

To compare our derived $f$(CO--dark H$_2$) from the GOT C+ survey we need models that cover all the possible parameters of density, metallicity, thermal pressure (density and temperature), and column density or mass.  Unfortunately none of the models in the literature covers such a wide range.  The work of \cite{Visser2009} has the advantage of considering a very large ensemble of clouds with many different physical parameters.  However, they only considered a metallicity appropriate to the local solar environment, whereas most of our sources are at smaller Galactic radii where the metallicity is larger. 

In summary we have shown that high spectral resolution surveys of [C\,{\sc ii}] emission can help illuminate the properties of the CO--dark H$_2$ distribution throughout the Galaxy and that clouds harbor a significant fraction of warm CO--dark H$_2$ gas.  Despite the limitations of a sparse survey and uncertainties in the density and temperature in the C$^+$ gas, with appropriate estimates of the temperature and density  we have been able to extract statistically significant information about the column density distribution and mass fraction of CO--dark H$_2$ in over 1800 clouds representative of the Galactic ISM. While the average column density of CO--dark H$_2$ clouds is about 10$^{21}$ cm$^{-2}$  there is considerable dispersion in the ensemble, consistent with the variations expected in models of diffuse molecular and transition clouds, but not with those for the densest, large column density clouds.  The fractional mass in CO--dark H$_2$ is largest in the diffuse clouds, $\sim$0.77 on average, reduced to $\sim$0.43 in the transition clouds, and least in the dense molecular clouds, $\sim$0.21. Finally, we find that the overall fraction of CO--dark H$_2$ increases with Galactic radius from $\sim$0.2 in the inner Galaxy to $\sim$0.5 at the solar radius.  
 

\begin{acknowledgements}

We thank the referee for a very careful reading of the manuscript and numerous suggestions for improving the presentation and clarifying the analysis and results.  We would also like to thank the staffs of the ESA Herschel Science Centre, NASA Herschel Science Center, and the HIFI Instrument Control Centre (ICC) for  their invaluable help with the data reduction routines. These included the standard HIPE routine, which is a joint development by the Herschel Science Ground Segment Consortium, consisting of ESA, the NASA Herschel Science Center, and the HIFI, PACS and SPIRE consortia, and a special purpose standing wave removal process developed at the HIFI ICC. 
This work was performed at the Jet Propulsion Laboratory, California Institute of Technology, under contract with the National Aeronautics and Space Administration.   
 
\end{acknowledgements}


\bibliographystyle{aa}
\bibliography{aa22406-13_Langer_refs}


\appendix

\section{Column Densities of H, C$^+$, and CO}
\label{sec:column-densities}

To determine the properties of the [C\,{\sc ii}]  clouds we need to calculate the column densities of the three main cloud layers as traced by: H\,{\sc i}, [C\,{\sc ii}], and CO along the line of sight of each spectral feature.  The CO and [C\,{\sc ii}] results are then combined to give the total column density of the H$_2$ gas, $N_{\rm tot}$(H$_2$)=$N$(CO--dark H$_2$)+$N$(CO--traced H$_2$).  All the column densities calculated here are beam averaged values. The results represent only one position in the cloud, however the assumption adopted here is that it represents on average the properties of the cloud.  We believe this assumption is reasonable for analyzing the properties of the CO--dark H$_2$ in the Galaxy because  GOT C+ detects so many spectral components it represents a statistical sampling of ISM clouds.    To calculate the column densities traced by each component we need to determine the excitation conditions and solve their radiative transfer equations.  There are uncertainties associated with each of these probes due to insufficient information about the densities and temperatures.  These uncertainties are not unique to the GOT C+  data, but have been present throughout the decades long history of using these ISM gas tracers.  Here we review the simple approach we adopted to calculate column densities efficiently for such a large ensemble of sources.  This approach, updated here, was first discussed in \cite{Langer2010} and \cite{Velusamy2010}. 


\subsection{H\,{\sc i} Column Density}

The column density of H\,{\sc i} can be estimated in the optically thin limit, where the brightness temperature is less than the kinetic temperature, as
\begin{equation}
N({\rm H\,I}) =1.82{\times}10^{18} I({\rm H\,I}) \,{\rm cm^{-2}} 
\end{equation}
\noindent where $I$(H\,{\sc i}) is the intensity in units of K \kmsnospace. In general the opacity of H\,{\sc i} clouds is small but difficult to determine on a cloud-by-cloud basis because it requires knowing the spin temperature, $T_{\rm s}$. However, in circumstances where the 21-cm line of H\,{\sc i} can be observed in absorption against background continuum sources it is possible to determine the spin temperature and H\,{\sc i} opacity, $\tau$(H\,{\sc i}).  
\cite{Kolpak2002}  observed 21-cm H\,{\sc i} lines in absorption against 54 compact extragalactic radio sources and derived the average H\,{\sc i} opacity, $<\tau >$, as a function of Galactocentric radius, $R_{\rm gal}$.  They find that it peaks in the molecular ring with a value $< \tau > \sim$1, but is much less outside this range.  However, they adopted a spin temperature of 50\,K, whereas 100\,K is more appropriate for the H\,{\sc i} clouds in our survey (see Paper I). Adopting a simple escape probability formula,  $\beta$ = (1-exp(-$\tau))/\tau$, the column density in the ring is underestimated by  $\sim$20$\%$ for $<\tau > \sim$0.5.  We adopt the approach in Paper I to include the H\,{\sc i} opacity, but it makes only a small correction for most sources. 


\subsection{[C\,{\sc ii}] Column Density}

To calculate the column density $N({\rm C^+})$ we adopt the approach in \cite{Langer2010} and \cite{Velusamy2012}, assuming optically thin emission.  A more detailed discussion of [C\,{\sc ii}] excitation, emission intensity, and column density can be found in \cite{Goldsmith2012}.  They find that for antenna temperatures up to 1$/$3 of the gas kinetic temperature, the antenna temperature is linearly proportional to the column density $N({\rm C^+})$, regardless of the optical depth of the transition -- a condition they refer to as ``effectively optically thin" or EOT.  
This EOT condition is satisfied for all of our [C\,{\sc ii}] components as the strongest line observed has a main beam 
antenna temperature T$_{\rm mb}$$\le$7 K, so we can safely use the optically thin radiative transfer equation.   
The column density in EOT for a gas layer dominated by either H or H$_2$, with no background radiation field (the CMB contribution is negligible), is derived from Equation (26) in \cite{Goldsmith2012},

\begin{equation}
N_j({\rm C^+}) = 2.9\times10^{15} [1+0.5e^{{\Delta}E/kT}(1+\frac{n_{\rm cr}(j)}{n(j)})]I_j([{\rm C\,II}]),
\end{equation}
    
\noindent where the energy of the excitation to the $^2P_{3/2}$ level, $\Delta E/k = 91.25$\,K, $j$= H or H$_2$ labels the gas layer and, thus collision partner, and $n(j)$ is the local density.  The intensity $I_j([{\rm CII}]) = \int{T_{\rm mb}([{\rm CII}])d{\rm v}}$ is in units of (K \kmsnospace), $N_{\rm j}({\rm C^+})$ ${\rm (cm^{-2})}$  is the corresponding column density of C$^+$, and $n_{\rm cr}(j)$ is the critical density.  The critical density is defined as, $n_{\rm cr} = A_{ul}/\left <\sigma {\rm v}\right>_{ul}$\, cm$^{-3}$, where $A_{ul}$ = 2.4$\times$10$^{-6}$ s$^{-1}$ is the Einstein spontaneous decay rate and $\left <\sigma {\rm v}\right>_{ul}$ is the collisional de-excitation rate coefficient, and $u$ and $l$ label the upper and lower states, respectively. For H collisions with C$^+$  we used the $\left<\sigma {\rm v}\right>_{ul}$ calculations of \cite{Barinovs2005}, yielding $n_{\rm cr}$(H) = 3.1$\times$10$^3$ cm$^{-3}$ at 100\,K.  The critical density for C$^+$ de-excitation by molecular hydrogen, H$_2$, is estimated to be twice that for atomic hydrogen, H\,{\sc i}  \citep[][]{Flower1988,Flower1990}.  We refer the reader to \cite{Goldsmith2012} for more details and fits to these collisional rate coefficients. 

For virtually all the sources in the GOT C+ survey the gas densities $n({\rm H})$ and $n(H_2)$ are much lower than the corresponding critical densites, $n_{\rm cr}(H)$ and $n_{\rm cr}(H_2)$, so that $n_{cr}(j)/n(j) >> 2$, which simplifies the bracket term in Equation A.2 as, 

\begin{equation}
[1+0.5e^{{\Delta}E/kT}(1+\frac{n_{\rm cr}(j)}{n(j)})] \simeq 0.5\frac{n_{\rm cr}(j)}{n(j)}e^{{\Delta}E/kT}.
\end{equation}

Therefore, to a very good approximation, we can write Equation A.2 as,

\begin{equation}
N_j({\rm C^+}) = 1.45\times10^{15} \frac{n_{\rm cr}(j)}{n(j)}e^{({\Delta}E/kT)}I_j([{\rm C\,II}])\,  {\rm (cm^{-2})}.
\end{equation}


\subsection{[C\,{\sc ii}] emission from H\,{\sc i} regions}

Given the column density of H\,{\sc i} in diffuse atomic clouds we can estimate the expected [C\,{\sc ii}] emission intensity, 
$I$([C\,{\sc ii}]) by first calculating the C$^+$ column density from $N_{{\rm H\,I}}({\rm C^+})=x_{\rm H}$(C$^+$)$N$(H\,{\sc i}), 
where the fractional abundance $x_{\rm H}$(C$^+$)=$n$(C$^+$)/$n({\rm H})$. In H\,{\sc i} clouds the gas phase
carbon is completely ionized so $x_{\rm H}$(C$^+$) is just the carbon metallicity.  Therefore, to a good approximation,

\begin{equation}
I_{\rm H\,I}([{\rm C\,II}]) = 6.9\times10^{-16} N_{\rm H\,I}({\rm C^+})\frac{n(H)}{n_{\rm cr}(H)}e^{-{\Delta}E/kT}\, (\rm K\, km\, s^{-1})
\end{equation}

\noindent where $T$ is the kinetic temperature of the H\,{\sc i}  gas. 

    
\subsection{CO--dark H$_2$ Column Density traced by [C\,{\sc ii}]}

In this paper we determine  the [C\,{\sc ii}] intensity arising from the CO--dark H$_2$ gas layer by subtracting the H\,{\sc i} component from the total [C\,{\sc ii}] intensity (the contributions from the WIM are neglected for the reasons discussed in Section 4.1),

\begin{equation}
 I_{\rm H_{2}} ([{\rm C\,II}]) =   I_{\rm total}([{\rm C\,II}]) -  I_{\rm HI}([{\rm C\,II}]).  
\end{equation}

\noindent For virtually all the GOT C+ sources  $n({\rm H}_2)$ is much less than the critical density, $n_{cr}({\rm H}_2)$, so that $n_{cr}({\rm H}_2)/n(\rm H_2)>>2$ and we can apply the approximation in Equation A.3 to yield,

\begin{equation}
N_{H_{2}}({\rm C^+}) = 1.45\times10^{15} \frac{n_{\rm cr}(H_2)}{n(H_2)}e^{({\Delta}E/kT)}I_{H_{2}}([{\rm CII}])\,  {\rm (cm^{-2})}.
\end{equation}

\noindent Then $N$(H$_{\rm 2}$)  traced by [C\,{\sc ii}] is given by

\begin{equation}
N_{[\rm CII]}({\rm H_2}) =  N({\rm C^+})/x_{\rm H_2}({\rm C^+}),
\end{equation}

\noindent where $x_{\rm H_2}$(C$^+$)= 2$x_H(\rm C^+)$  The total H$_2$ column density is then given by, $N_{\rm tot}$(H$_2$)=$N$(CO--dark H$_2$)+$N$(CO--traced H$_2$) = $N_{[\rm CII]}({\rm H_2})$ +$N_{\rm CO}(H_2)$ .

 
\subsection{Column density of H$_2$ from CO}

In Paper I we used both $^{12}$CO and $^{13}$CO as tracers of Galactic molecular gas and discussed the regimes for using these two tracers. In this paper we use just the $^{12}$CO as a tracer of molecular H$_2$ in shielded regions in order to have a consistent and uniform estimate of the column density, and also because the number of clouds with $^{13}$CO is smaller than that with $^{12}$CO (all clouds in which we detect $^{13}$CO are also detected in $^{12}$CO).   However, as discussed above, we use the detection of $^{13}$CO (and C$^{18}$O) to label the type of cloud under study. 

To derive the column density of H$_2$ traced by $^{12}$CO we utilize the result established between $N$(H$_2$) and the intensity of $^{12}$CO(1$\to$0), $N_{\rm {CO}}({\rm H_2}) = X_{\rm {CO}}I({\rm ^{12}CO})$\,cm$^{-2}$, where $X_{\rm {CO}}$ is the conversion factor, which is based on the observation that there is a relationship among size-linewidth, virial mass, and CO intensity first developed by \cite{Solomon1987} (see also the detailed review by \citealt{Bolatto2013}).  This approach works in part because $^{12}$CO becomes optically thick at a short penetration depth into the cloud once CO can be efficiently produced and it measures the surface properties of a cloud in virial equilibrium, and thus traces the mass inside the $\tau$ =1 to 2 surface.  As discussed by \cite{Bolatto2013} the conversion factor will depend somewhat on metallicity because the transition from C$^+$ to CO depends on the dust shielding and CO self-shielding, which are clearly less if the metallicity is lower.  Indeed studies of other galaxies provide evidence of this dependence but there is much scatter and uncertainty from source-to-source.  In Paper I we derived $X_{\rm {CO}}$ from our estimation of the H$_2$ column density only from the $^{12}$CO and $^{13}$CO emissivities (that is excluding the CO--dark H$_2$ component) and found evidence for a radial gradient similar to that measured in local galaxies (\citealt{Wilson1995}).   The majority of our $^{12}$CO sources are located between 4 and 7 kpc and the change in $X_{\rm {CO}}$ suggested in Paper I and \cite{Wilson1995} is about 25$\%$, less than the $\pm$30$\%$ uncertainty quoted by \cite{Bolatto2013} for $X_{\rm CO}$ in the Milky Way.  Given the uncertainties in $X_{\rm CO}$ and that the main focus of our paper is on the CO--dark H$_2$ gas we take a simple approach and adopt a constant value from  \cite{Dame2001}, $X_{\rm {CO}}$=1.8$\times$10$^{20}$ (K \kmsnospace)$^{-1}$.  

Finally, we compare $^{12}$CO and $^{13}$CO to see how well they track each other in those clouds where both are detected.  In principle, if optically thin (or nearly optically thin) $^{13}$CO measures the column density directly and the optically thick $^{12}$CO measures it indirectly, then these should be correlated.   In Figure~\ref{fig:fig_A1_13COvs12CO}  we plot the integrated intensities of $^{13}$CO versus those of $^{12}$CO for the 734 sources where we detect both isotopologues. Models of CO isotopic chemistry predict that $^{13}$CO becomes abundant as soon as does $^{12}$CO because the isotopic exchange reaction, $^{13}{\rm C}^{+}$ +  $^{12}{\rm CO} \rightarrow$  $^{13}{\rm CO}$ +  $^{12}{\rm C}^{+}$, is fast. So wherever there is abundant $^{12}$CO we expect  $^{13}$CO should be present.   The threshold for the onset of $^{13}$CO seen in this figure is largely due to the difficulty of detecting the less abundant $^{13}$CO in the weaker $^{12}$CO sources, and not because $^{13}$CO is absent.  There is also a large degree of scatter that is likely due to sampling a variety of cloud conditions.  Nonetheless, we see an excellent correlation of  $I$($^{13}$CO) with $I$($^{12}$CO), confirming that, on average, $I$($^{12}$CO) is a good linear tracer of $N$(H$_2$) in the CO shielded regions of clouds for $I$($^{12}$CO)$>$5 K \kmsnospace. A linear fit to the CO intensities plotted in Figure~\ref{fig:fig_A1_13COvs12CO} yields, $I$($^{12}$CO)/$I$($^{13}$CO)$\simeq$7 with a correlation coefficient of 0.84; in agreement with the range for $I$($^{12}$CO)/$I$($^{13}$CO) of 4 to 8 found in surveys of the Galactic ring in $^{12}$CO and $^{13}$CO  \citep[c.f.][]{Sanders1993}. Converting these intensities to abundance ratios is difficult due to the opacity of the abundant $^{12}$CO isotopologue.  The median I($^{12}$CO) in the GOT C+ sources is $\sim$12 K \kmsnospace, and the corresponding opacity for the J=1$\rightarrow$0 line is of order a few, assuming typical dense cloud densities and temperatures.


\begin{figure}
  \centering
 \includegraphics[width=7.5cm]{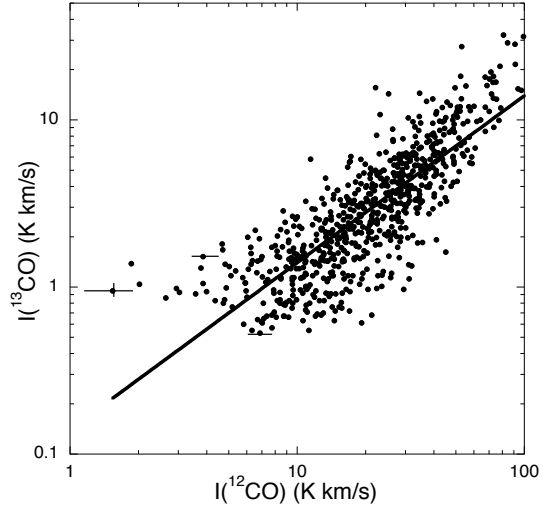}
      \caption{Scatter plot of $I$($^{13}$CO) versus $I$($^{12}$CO) for the 734 sources detected in $^{13}$CO.  The solid line is a linear fit to the data points in the figure and indicates that $I$($^{12}$CO) is, on average, a good linear tracer of H$_2$ column density for $I$($^{12}$CO)$>$5 K \kmsnospace. The 1-$\sigma$ noise is indicated on two representative low intensity points; on the scale of this figure it can be seen that the signal-to-noise ratio is very high except for a few of the weakest $^{12}$CO points.  The detection limits are given in Table~\ref{tab:sensitivity}.}
              \label{fig:fig_A1_13COvs12CO}
    \end{figure}
   

\end{document}